\documentclass[12pt]{article}
\usepackage[left=2cm,right=2cm,top=2.8cm,bottom=2.8cm]{geometry}

\usepackage[colorlinks=true,linkcolor=black,citecolor=black,urlcolor=blue,filecolor=black]{hyperref}

\usepackage{tensor}
\usepackage{amsmath}
\usepackage{amsfonts}
\usepackage{amssymb,amsthm,mathabx}
\usepackage{color}
\usepackage{ytableau}
\usepackage{enumerate}
\usepackage{cite}

\numberwithin{equation}{section}

\def\d{\delta}

\newcommand{\half}{\frac{1}{2}}
\renewcommand{\d}{\partial}

\newcommand{\ffrac}[2]{\raisebox{.5pt}
  {\footnotesize$\displaystyle\frac{#1}{#2}$}\kern1pt}

\newcommand{\ddl}[2]{\ffrac{\d #1}{\d #2}}

\def\cA{{\cal A}}

\def\cC{{\cal C}}

\def\cI{{\cal I}}

\def\cL{{\cal L}}

\def\be{\begin{equation}}
\def\ee{\end{equation}}
\def\bea{\begin{eqnarray}}
\def\eea{\end{eqnarray}}
\def\ba{\begin{array}}
\def\ea{\end{array}}

\def\12{\frac{1}{2}}

\ytableausetup{mathmode, boxsize=1.4em}
\newcommand{\YT}[1]{\begin{ytableau}#1\end{ytableau}}

\begin{document}

\vspace{30pt}

\begin{center}

  {\Large\sc A note on local BRST cohomology of Yang-Mills type theories with free
    abelian factors}

\vspace{25pt} {\sc Glenn Barnich${}^{\, a}$ and Nicolas
  Boulanger${}^{\, a, b}$}

\vspace{10pt} {${}^a$\sl\small Universit\'e libre de Bruxelles and
  International Solvay Institutes, Campus Plaine CP231, B-1050
  Brussels, Belgium \vspace{10pt}

  ${}^b$\sl \small Groupe de M\'ecanique et Gravitation, Physique
  Th\'eorique et Math\'ematique,
  Universit\'e de Mons -- UMONS, 20 Place du Parc, B-7000 Mons, Belgium\\
  \vspace{10pt}}

\vspace{40pt} {\sc\large Abstract}
\end{center}

\noindent

We extend previous work on antifield dependent local BRST cohomology
for matter coupled gauge theories of Yang-Mills type to the case of
gauge groups that involve free abelian factors. More precisely, we
first investigate in a model independent way how the dynamics enters
the computation of the cohomology for a general class of Lagrangians
in general spacetime dimensions. We then discuss explicit solutions in
the case of specific models. Our analysis has implications for the
structure of characteristic cohomology and for consistent deformations
of the classical models, as well as for divergences/counterterms and
for gauge anomalies that may appear during perturbative quantization.

\newpage

\begin{footnotesize}
\setcounter{tocdepth}{3}
\tableofcontents
\end{footnotesize}

\newpage

\section{Introduction}\label{sec:intro}

The use of systematic algebraic methods has proved extremely useful in
the context of renormalization of vector gauge models
\cite{Becchi:1975nq}. A subsequent reformulation in terms of the
effective action \cite{Zinn-Justin:1974mc} streamlines the analysis
(see \cite{Piguet:1995er} for a review) and has paved the way for
generalizations to generic gauge systems \cite{Batalin:1981jr}. In
this context, a detailed understanding of how the antifield dependent
BRST differential encodes the information about gauge symmetries and
dynamics \cite{Fisch:1990rp} is a pre-requisite for the computation of
the relevant cohomologies.

It turns out that it is the antifield dependent BRST cohomology
computed in the space of local functionals that is needed for
perturbative quantization and renormalization of gauge models. Indeed,
even when no power counting restrictions are imposed
\cite{Gomis:1996jp}, these cohomology classes control in a gauge
independent way both anomalies (in ghost number $1$) and non-trivial
divergences/counterterms (in ghost number $0$) that cannot be absorbed
by a field-antifield redefinition, including a change of gauge
\cite{Voronov:1982ph}.

The antifield dependent part of the cohomology depends on the dynamics
of the model through (generalized) conservation laws and global
symmetries and their interplay with the gauge structure
\cite{Barnich:1994db,Barnich:1994mt}. In the case of semi-simple gauge
groups, this dependence can only occur through gauge invariant
conserved currents, and is absent in ghost numbers $0$ and $1\,$.  As
a consequence, there is no non-trivial renormalization of the BRST
symmetry itself since it is encoded in the antifield dependent part of
the master action.

If there are abelian factors, the situation becomes more involved (see
\cite{Bandelloni:1978ke,Bandelloni:1978kf} for an early analysis) and
additional antifield dependent cohomology may appear, both in ghost
number $0$ and in ghost number $1$. If some of these abelian factors
are free in the sense that the matter fields do not transform under
the associated gauge transformations, the antifield dependence becomes
even more complicated. From the point of view of renormalization of
vector gauge theories, such models are usually not considered even
though they should in case the model is not completely free. For
instance, in the bosonic sector of supergravity models with gravity
switched off, there typically are couplings of the curvatures of the
vector gauge fields to the scalar fields.

Besides non-trivial divergences/counterterms, antifield dependent BRST
cohomology in ghost number $0$ also controls non-trivial deformations,
so-called gaugings. The problem of systematically finding all such
gaugings has recently attracted renewed interest in the context of
supergravities (see \cite{Trigiante:2016mnt} for a review). In this
case, free abelian factors feature prominently from the very
beginning. It is this question that consitutes the main motivation
behind the present analysis (see \cite{Henneaux:2017kbx} for more
details and the companion paper \cite{Barnich:2017nty} for detailed
results in 4 dimensions with only free abelian factors).

It is thus of interest to have as complete results as possible on
these cohomologies. Besides the illustration of how the complicated
antifield dependence of the cohomology leads to standard non-abelian
Yang-Mills models when starting from pure, free abelian vector fields,
section 12 of reference \cite{Barnich:1994mt} outlines how to proceed
in more general cases with non trivial couplings to matter
fields. More precisely, the equations that are affected by the
modified dynamics have been identified, and a case by case analysis
has been suggested.

The purpose of the present paper is to formulate in detail and in a
model independent way the dynamical equations to be solved and how
they appear as building blocks in the charcaterization of the
cohomology, before working out explicit results in specific models.

In the case without free abelian vector fields, this had already been
done in the review \cite{Barnich:2000zw} by following a different
strategy than the one adopted in \cite{Barnich:1994db,Barnich:1994mt}:
instead of computing the cohomology by attacking the cocycle condition
in top form degree at highest antifield number, one starts the
analysis from the bottom of the descent equations. The advantages of
this modified strategy are twofold: (i) The equivariant characteristic
cohomology can be determined together with the antifield dependent
local BRST cohomology in a unified and streamlined proof, and (ii) one
may forgo the technical assumption of normality. More precisely, for
the approach of \cite{Barnich:1994db,Barnich:1994mt} to be viable, one
needs to prove that various expansions have bounded antifield
number. This can be done by suitable assumptions on the number of
derivatives in the interactions. Effective theories are however not
normal in this sense. In this case, one needs to consider the space of
formal power series in the couplings constants, the fields, antifields
and their derivatives, so that the antifield number may be unbounded.
The approach of \cite{Barnich:2000zw} allows one to cover both
situations simultaneously. That is why the terminology ``normal
theories'' was extended so as to include effective theories with
unbounded expansions and no additional assumptions on derivatives (see
sections 5.3, 6.4.3 and 7.3 of \cite{Barnich:2000zw} for details).

The present paper completes the results obtained in
\cite{Barnich:2000zw} to the case where there are free abelian gauge
fields. The analysis is valid in all spacetime dimensions greater than
or equal to three. In the present case, the dynamics enters the
problem non-trivially already at form degree $n-2$ and involves two
additional building blocks.

In the next section, we summarize the ingredients of
\cite{Barnich:2000zw} needed for a self-contained presentation of the
main results. In addition, we provide a new characterisation in terms
of suitable Young tableaux of antifield independent local BRST
cohomology
\cite{DuboisViolette:1985hc,DuboisViolette:1985jb,Brandt:1989rd,%
  Brandt:1990gy,Dubois-Violette:1992ye} in the case where there are
only abelian factors. Our analysis extends the considerations in
\cite{DuboisViolette:1985cj} and allows one to make contact with the
topological gaugings of \cite{deWit:1987ph}.

Section~\ref{sec:theorem} is devoted to the statement of the main
theorem on the local BRST cohomology $H(s|d)$ and its implications for
characteristic cohomology and for infinitesimal gaugings. In section
\ref{sec:appl-known-models}, the general results are illustrated in
the cases of vector and vector-scalar models in various
dimensions. Finally, appendix \ref{sec:notations} contains our
conventions regarding forms, while appendix \ref{sec:proof} contains
the recursive proof of the main theorem.

\section{BRST-BV differential for Yang-Mills type theories}
\label{sec:bv-differential-yang}

\paragraph{Basic definitions.} 
The generic Yang-Mills-type gauge algebra $\mathfrak{g}\,$ that we
consider is a direct sum of several abelian and simple Lie algebras.
In a basis $t_I$ of $\mathfrak{g}\,$ we have
$[t_I,t_J]=t_K\,f^K{}_{IJ}\,$,
$I,J,K \in \{1,\ldots,\rm{dim}(\mathfrak{g})\}\,$.  We denote by
${\cal I}$ the space of $\mathfrak g\,$-invariant local functions in
$x^\mu, dx^{\mu}, F^I_{\mu\nu}, D_\rho F^I_{\mu\nu}, \dots$,
$\psi^i, D_\mu\psi^i,\dots\,$, where
\begin{align}
 & F^I_{\mu\nu}:=2\partial_{[\mu}A^I_{\nu]}
 +\,f^I{}_{JK}\,A^J_{\mu}A^K_{\nu}\;,\quad  A^I:=dx^{\mu}\,A^I_\mu\;,\\
 & D_\rho F^I_{\mu\nu} := \partial_\rho F^I_{\mu\nu} + \,f^I{}_{JK}
 A_{\rho}^J\,F^K_{\mu\nu}\;,\quad 
 D_\mu\psi^i := \partial_{\mu}\psi^i + \, T_I{}^i{}_j \,A^I_{\mu}\,\psi^j
\end{align}
are, respectively, the components of the field strengths, the gauge
one-forms, the $\mathfrak{g}\,$-covariant derivatives of the field
strengths and of the (scalar or spinorial) matter fields $\psi^i$.
The latter transform in a dim$_\psi\,$-dimensional representation of
$\mathfrak{g}$ with matrices
$\{T_I{}^i{}_j\}_{I=1, \ldots, \rm{dim}(\mathfrak{g})}\,$,
$i,j\in \{1,\ldots, \rm{dim}_\psi\}\,$.

The solution of the BV master equation for Yang-Mills theories is given by
\begin{equation}
  \label{eq:32}
  S^{(0)}=\int d^nx [\cL +A^{*\mu}_I D_\mu C^I+\half
  C^*_I {f^I}_{JK}C^JC^K  + C^I{{T_I}^i}_j\psi^i\psi^*_i ]\,.
\end{equation}
When introducing the antifield number \emph{antifd}, which assigns
antifield numbers $0$ to the fields and ghosts, $1$ to
$A^{*\mu}_I,\psi^*_i$ and $2$ to $C^*_I$, the associated BRST
differential decomposes into $s = (S^{(0)},\cdot)= \delta + \gamma\,$,
where the longitudinal differential $\gamma$ preserves the antifield
number and the Koszul-Tate differential $\delta$ decreases the
antifield number by $1$. For more details on the Koszul-Tate
differential and its relevance for the conservation laws and rigid
symmetries of a local theory, see \cite{Henneaux:1992ig} as well as
sections 5 and 6 of \cite{Barnich:2000zw}.

\paragraph{Small algebra and descent of equations.} 
\label{sec:refin-basis-abel}
The small algebra $\cal B$ is defined to be the algebra of polynomials
in the undifferentiated ghosts, the gauge field 1-forms and their
exterior derivatives.  In the review \cite{Barnich:2000zw}, following
the orginal work \cite{DuboisViolette:1985jb}, a basis of
$H^{*,p}(s,\cal B)$, adapted to the computation of
$H^{*,*}(s|d),\cal B)$, was constructed in terms of forms $M$ and $N$
satisfying
\begin{align}
    s B^p = - d(B^{p-1} + M^{p-1})\;,\quad 
    d B^p = -s(B^{p+1} + b^{p+1}) + N^{p+1}\;,
    \label{11.37}
\end{align}
for suitably defined $B^{p-1}$, $B^{p+1}$ and $b^{p+1}$ in
${\cal B}\,$. As a consequence, a basis of $H^{*,*}(s|d,\cal B)$ is
provided by the $B$'s and the $M$'s and $1$. We will not reproduce
explicit expressions here but refer instead to \cite{Barnich:2000zw},
sections 10.5-10.7.

If the gauge group $G$ contains only $N$ abelian factors, we now
provide an alternative adapted basis, that is more suitable
for our purpose below. Indeed, a basis of $H^*(s,\cal B)$ is provided
by $1$ together with all the homogeneous polynomials
$p_{b,k} = t_{J(b);I[k]}\,F^{J(b)}\,C^{I[k]}\,$ of degree $b$ in the
$F^J$'s and of degree $k$ in the $C^I$'s. Here, we have used the
notation $F^{J(b)}=F^{J_1}\ldots F^{J_b}\,$, and similarly for
$C^{I[k]}=C^{I_1}\ldots C^{I_k}\,$. The alternative basis
of generators of $H^*(s,\cal B)$, that is adapted to the computation
of $H^{*,*}(s|d,\cal B)$ in the purely abelian case, is given by
\begin{align}
  \{ M_{b,k} \} & = 
  \{ \, \lambda_{J(b)I,I[k-1]}\,F^{J_1}\ldots F^{J_b}\,C^{I_1}\ldots 
  C^{I_k} \,\}\;, \qquad k > 0\;, b\geqslant 0\;, 
\label{B.38} \\
  \{ N_{b+1,k-1} \} & = 
  \{ \, \tfrac{b+k}{b+1}\;\lambda_{J(b+1),I[k-1]}\,F^{J_1}\ldots F^{J_{b+1}}\,C^{I_1}\ldots 
  C^{I_{k-1}} \,\}\;, \qquad k > 0\;, b\geqslant 0\;, \quad 
\label{B.38b}
\end{align}
in terms of constant tensors
$\lambda_{J_1\ldots J_{b+1},I_1\ldots I_{k-1}}$ satisfying
\begin{align}
& \lambda_{J_1\ldots J_{b+1},I_1\ldots I_{k-1}} = 
\lambda_{(J_1\ldots J_{b+1}),I_1\ldots I_{k-1}} = 
\lambda_{J_1\ldots J_{b+1},[I_1\ldots I_{k-1}]} \;,
\\
\quad & 
\lambda_{(J_1\ldots J_{b+1},J_{b+2})I_2\ldots I_{k-1}}=0\;.
\label{eq:lamsym}
\end{align}
In other words, the constants $\lambda_{J_1\ldots J_{b+1},I_1\ldots I_{k-1}}$ 
transform in the $GL(N)$ Young tableau associated
with an upper row of length $b+1\,$ whose boxes are filled with the
indices $\{J_1,\ldots, J_{b+1}\}\,$, while all the lower rows are of
length one and filled with the indices $\{I_1,\ldots , I_{k-1}\}\,$.
Based on these properties, one can check that:

(i) Together with $1$, these $M$'s and $N$'s do define a basis of
$H^*(s,\cal B)$.  Indeed, any polynomial
$p_{b,k} = t_{J(b);I[k]}\,F^{J(b)}\,C^{I[k]}$ is represented by a
$GL(N)$ reducible tensor $t_{J(b);I[k]}\,$.  The latter decomposes in
the direct sum of irreducible tensors $\lambda_{J(b+1),I[k-1]}$ and
$\lambda_{J(b),I[k]}\,$. For example, the decomposition of the
reducible tensor $t_{J(2);I[3]}\equiv t_{J_1J_2;I_1I_2I_3}$ into
$\lambda^{(M)}_{J_1 J_2 I_1, I_2 I_3}$ and
$\lambda^{(N)}_{J_1 J_2 , I_1 I_2 I_3}$ can be depicted as follows:
\begin{align}
    \YT{J_1 & J_2} ~\otimes~ \YT{I_1 \\I_2\\I_3} & ~\sim~ 
    \YT{I_1 & J_1 & J_2 \\I_2\\I_3} ~~\;\oplus ~~\YT{J_1 & J_2 \\I_1\\I_2\\I_3}
    \\
    t_{J_1 J_2 ; I_1 I_2 I_3}  & ~=~ \lambda^{(M)}_{J_1 J_2 [I_1, I_2 I_3]} ~+~ 
    \lambda^{(N)}_{J_1 J_2 , I_1 I_2 I_3}\;.
\end{align}

(ii) The following descent equations hold
\begin{align}
s  ( \tfrac{1}{2}\,A^{K}A^{L}\,\partial_{L}\partial_{K}\,M_{b,k} ) 
  + {\rm d} (A^{K}\partial_{K}\,M_{b,k} )
   &  = N_{b+1,k-1} \;, \nonumber \\
s( A^{K}\partial_{K}\,M_{b,k} ) + {\rm d} M_{b,k}  
    &  = 0 \;, \nonumber \\
s M_{b,k} & = 0 \;. \label{abeliandescent} 
\end{align}
Following the arguments given in \cite{Barnich:2000zw}, it then
follows that a basis of $H^{*,*}(s|d, \cal B)$ is given by $1$,
$M_{b,k}$ and $ A^{K}\partial_{K}\,M_{b,k}$.

One may now compute the dimensions of $H^{*,*}(s\vert d,\cal B)\,$ and
compare to those given in \cite{DuboisViolette:1985cj}. From equations
\eqref{B.38}, \eqref{B.38b}, \eqref{abeliandescent} and the theorems
9.1 and 9.2 of \cite{Barnich:2000zw},
$H^{q,2p}(s\vert d,\cal B)\,$ and $H^{q-1,2p+1}(s\vert d,\cal B)\,$
are isomorphic. The dimension of $H^{q,2p}(s\vert d,\cal B)\,$ is
simply given by the independent components of the tensor
$\lambda_{J_1\ldots J_{p+1},I_1\ldots I_{q-1}}\,$ appearing in
$\lambda_{J_1\ldots J_{p} I_{1},I_2\ldots I_{q}} \,F^{J_{1}}\ldots
F^{J_{p}}\,C^{I_{1}}\ldots C^{I_{q}}\,$.  In other words, $h^{p,q}:=$
dim$(H^{q,2p}(s\vert d,\cal B))\,$ is equal to the dimension of the
$GL(N)$\, Young tableau
\begin{align} 
    \YT{I_1 & J_1 & \ldots  & J_p \\I_2\\ \vdots \\I_{q} }\;,
\end{align}
which is \cite{hamermesh89}
\begin{align}
h^{p,q} = \frac{(N+p)!}{(N-p)!\,p!\,(q-1)!\,(p+q)}\;\;.
\label{dimhookGLN}
\end{align}
In \cite{DuboisViolette:1985cj}, the dimensions $h^{p,q}$ were
encoded in the generating function
\begin{align}
h(x,y) = \sum_{p+q\geqslant 1} h^{p,q}\,x^{p}\,y^{q} =
  \frac{y}{1-x}\;\sum_{r=0}^{N-1}\Big(\frac{1+y}{1-x}\Big)^{r}\;. 
\end{align} 
To show that these dimensions indeed agree, we first expand the
geometric series, 
\begin{align}
h(x,y) = y \sum_{m=0}^{\infty}x^{m}+y(1+y)
  \sum_{m=0}^{\infty}(m+1)\,x^{m} +y(1+y)^{2}
  \sum_{m=0}^{\infty}\tfrac{(m+1)(m+2)}{2!}\,x^{m} 
\nonumber\\
+\ldots + y(1+y)^{N-1} \sum_{m=0}^{\infty}\tfrac{(m+1)(m+2)\ldots
  (m+N-1)}{(N-1)!}\,x^{m}\;,
\nonumber
\end{align}
then expand $(1+y)^{p}$ and collect the coefficient of $x^{p}y^{q}$:
\begin{align}
h^{p,q} = {}_{p+q-1}C_{q-1}\Big[ 1 + (p+q) +
  \tfrac{(p+q)(p+q+1)}{2!}\,+\ldots+
  \tfrac{(p+q)\ldots(p+q+\tilde{N}-1)}{\tilde{N}!}\,  
\Big]\;,\nonumber
\end{align}
where ${}_{n}C_{k}:=\frac{n!}{(n-k)! k!}\,$ are the binomial
coefficients and $\tilde{N}:=N-q\geqslant 0\,$.  If we now
view $h^{p,q}$ as a function of the non-negative integer
$\tilde{N}$ and define
$\sigma({\tilde{N}}):=h^{p,q}/{}_{p+q-1}C_{q-1}$, we can determine it
by induction using 
\begin{align}
\sigma({\tilde{N}+1}) = \sigma({\tilde{N}}) +
  {}_{\tilde{N}+p+q}C_{\tilde{N}+1}\;,\quad \sigma(0) = 1\;.  
\end{align}
It is then easy to see that the solution is
$\sigma(\tilde{N}) = {}_{\tilde{N}+p+q}C_{\tilde{N}}\,$. This gives
$h^{p,q}=\frac{(\tilde{N}+p+q)!}{\tilde{N}! (p+q) p!(q-1)!}\,$ and
then \eqref{dimhookGLN} when using the definition of $\tilde{N}\,$.

\section{The main theorem}
\label{sec:theorem}

In this section, theorem 11.1 of \cite{Barnich:2000zw} is extended so
as to include free abelian factors. As a consequence, there now exists
non-trivial characteristic cohomology already in form degree $n-2\,$,
$H^{n-2}_{\rm char}(d )\cong H^{-2,n}(s \vert d)$
\cite{Barnich:1994db,Barnich:1994mt}, which in turn considerably
enriches $H^{g,n}(s|d)$ in ghost numbers $g$ greater than $-2$.

The plan of this section is as follows. In subsection
\ref{subsec:assumptions}, we spell out the assumptions underlying the
theorem. In subsection \ref{sec:comm-char-cohom}, we start by
classifying elements of characteristic cohomology
$H^{n-k}_{\rm char}(d)$, for $k=1,2$ according to the shortest length
of descent of the associated elements of $H^{-k,n}(s\vert d)\,$. In
subsection \ref{sec:ingredients}, we list the main building blocks
needed for the theorem. The theorem itself is stated in subsection
\ref{sec:theorem-11.1-bis} and proved by induction in appendix
\ref{sec:proof}.  As a first corollary to the theorem, we provide a
complete discussion of non-covariantizable characteristic cohomology
and its relation to equivariant characteristic cohomology in
subsections \ref{sec:cons-char-cohom} and
\ref{sec:cons-char-cohombis}. Finally, we discuss implications for
infinitesimal deformations in subsection \ref{sec:interpr-results}.

\subsection{Assumptions}
\label{subsec:assumptions}

Our assumptions are as follows:

\begin{enumerate}[a)]

\item There \emph{are} free abelian gauge fields: the gauge group
  contains abelian factors such that the corresponding gauge
  transformations on the abelian gauge fields $\{A^{I_{fa}}_\mu\}$ and
  on the matter fields $\{\psi^i\}$ are
  $\delta_{\epsilon^{I_{fa}}}
  A^{I_{fa}}_\mu=\partial_\mu\epsilon^{I_{fa}}$ and
  $\delta_{\epsilon^{I_{fa}}}\psi^i = 0\,$. In other words, the matter
  fields are uncharged under the free abelian factors.  This has the
  consequence that the action of the BRST differential $s$ on 
  the  $C^*_{I_{fa}}$ gives no contributions in
  the antifields $\psi^*_i$ and only produces
  $\delta C^*_{I_{fa}} = - \partial_\mu A^{*\mu}_{I_{fa}}\,$;

\item The spacetime dimension is strictly greater than two, $n>2\,$; 

\item We consider only Lagrangians $\cL$ obeying the regularity
  conditions spelled out in the section 5.1.3 of \cite{Barnich:2000zw}
  and defining normal theories, as explained in the section 6.4.3 of
  the same reference.  The left-hand sides of the field equations
  $\frac{\delta {\cal L}}{\delta \psi^i}\approx 0$ and
  $\frac{\delta {\cal L}}{\delta A^I_\mu}\approx 0$ are assumed to be
  gauge covariant;

\item We assume that there is no nontrivial topology: 
  the cohomology of $d$ is trivial except for the constants 
  and the inequivalent Lagrangians\footnote{It would be 
    interesting to relax this assumption in order to include
    magnetic-type charges.};

\item There are only matter and vector fields and the only gauge
  invariance is of Yang-Mills type with a reductive\footnote{This
    assumption could also be relaxed if needed. In the non-reductive
    case, functions in the covariant derivatives of the field
    strengths and the undifferentiated ghosts are not necessarily
    separately $\mathfrak g$ invariant, see
    e.g. \cite{Barkallil:2002fp} for a discussion.} gauge algebra
  $\mathfrak g\,$.

\end{enumerate}

\subsection{Comments on characteristic cohomology}
\label{sec:comm-char-cohom}

Under our assumptions, it is shown in
\cite{Barnich:1994db,Barnich:1994mt,Barnich:2000zw} that
characteristic cohomology in degree $n-2$ is isomorphic to
$H^{-2,n}(s|d)\,$, a basis being given by $d^nx C^*_{I_{fa}}$ where
the index $I_{fa}$ is restricted to run over the free abelian
factors. The associated descent equations start with
\begin{align}
s\, d^nx\,C^*_{I_{fa}} + d \star A^*_{I_{fa}} & =0 \;,
\\
s \star A^*_{I_{fa}}+d k^{n-2}_{I_{fa}} & =0\;.    
\end{align}
In this case, characteristic cohomology in degree $n-2$ is represented
by the $k^{n-2}_{I_{fa}}$.

By applying $s$ to the last equation, one then gets 
 $d s k^{n-2}_{I_{fa}} =0 \,$,   
which implies
\begin{align}
s k^{n-2}_{I_{fa}}+d r^{n-3}_{I_{fa}}=0\;.
\end{align}
From the general analysis of descent equations (see sections
9.2 and 9.3 of \cite{Barnich:2000zw}), it follows that
the shortest length of the descent, also called depth below, allows
one to distinguish elements of $H^{-2,n}(s|d)\,$.  More precisely, we
split the elements into those for which the descent stops at form
degree $n-2$ and ghost number $0\,$, and those for which
$r^{n-3}_{I_{fab}}$ cannot be absorbed by allowed redefinitions.  A
basis for the former free abelian factors is denoted by $d^nx C^*_a$
and for the latter by $d^nx C^*_{\cA}\,$.  Since $s k^{n-2}_a=0$ and
$k^{n-2}_a$ is of ghost number $0$ and does not involve antifields, it
follows that $k^{n-2}_a\in \cI\,$, while $k^{n-2}_{\cA}$ cannot belong
to $\cI\,$.  For the latter class, one has
\begin{align}
s\, d^nx\,C^*_{\cal A} + d \star A^*_{\cal A} & =0 \;,
 \nonumber \\
s \star A^*_{\cal A} + d k^{n-2}_{\cal A} & =0\;,
\label{noncovariantcharactn-2a}\\
s k^{n-2}_{\cal A} + d r^{n-3}_{\cal A} & = 0\;, 
\quad 
r^{n-3}_{\cal A} \neq s p^{n-3}_{\cal A} + d q^{n-4}_{\cal A}
                                          \;. \nonumber 
\end{align}

For characteristic cohomology in degree $n-1$, we consider the descent
equations starting with $s \omega^{-1,n}+d \omega^{0,n-1}=0\,$. In this
case, characteristic cohomology in degree $n-1$ is represented by the
antifield independent part of $\omega^{0,n-1}$. For the shortest
descent, one has $s \omega^{0,n-1}=0$ and a basis for $(n-1)\,$-forms
at the bottom of the descent is denoted by $j_\Delta\,$.

Finally, the equation $P^n_A(F)\approx dI^{n-1}_A$ can be shown to
imply $P^n_A(F)=dI^{n-1}_A+s K_A\,$ where the $K_A$ depend
linearly on undifferentiated antifields and $\gamma K_A=0$. More
generally, if $\cI\ni I^n\approx 0\,$, then
$I^n+d(\delta R^{n-1})=sK\,$, where $K$ can be assumed to depend
linearly on undifferentiated antifields only and
$\delta R^{n-1}\in \cI\,$ --- see the proof in section 12.5 of
\cite{Barnich:2000zw}. 

\subsection{Ingredients}
\label{sec:ingredients}

The theorem rests on the following ``elementary'' descent equations:
\begin{enumerate} 

\item For
  all non-trivial solutions to the descent equations in $\cal B\,$, we
  have equation \eqref{11.37}: $\,sB^p+d(B^{p-1}+M^{p-1})=0\,$,
  $dB^p=-s( B^{p+1}+b^{p+1})+N^{p+1}\,$.  In particular, for
  $\Theta_\alpha$ a basis of the $s$ cohomology in the space of
  undifferentiated ghosts, we have
\begin{align}
    s[\Theta_\alpha]^2+d[\Theta_\alpha]^1 & =N^2_\alpha
          =P_\alpha^{2\beta}(F)\Theta_\beta\;,\nonumber \\
    s[\Theta_\alpha]^1+d\Theta_\alpha & =0\;, \label{C.8}\\
    s\Theta_\alpha & = 0\;.\nonumber
  \end{align}

 Note that the obstruction $N^2_\alpha$ exists
 only if $\Theta_\alpha$ involves an abelian factor.
 
\item A basis for gauge-invariant characteristic cohomology in form
  degree $n-2$ is given by the
  $k^{n-2}_a \equiv -d^{n-2}x_{\mu\nu}k_a^{\mu\nu}\in \cI\,$. They
  satisfy $d k^{n-2}_a\approx 0\,$, or more precisely,
  $\frac{\delta \cL}{\delta A^a_\mu}=\d_\nu k^{\nu\mu}_a\,$.  The
  $k^{n-2}_a$'s constitute a basis in the sense that
  $dI^{n-2}\approx 0\;\Rightarrow\; I^{n-2}\approx \lambda^a
  k_a+d\omega^{n-3}$, and
  $\lambda^a k_a\approx d\eta^{n-3}\;\Rightarrow\; \lambda^a=0$. They
 are related to the $n\,$-forms $d^nx\,C^*_a\,$ through the chain:
\begin{align}
s\, d^nx\,C^*_a + d \star A^*_a & =0 \;,\nonumber
\\
s \star A^*_a+d k^{n-2}_a & =0\;,
\label{C.11}\\
s k^{n-2}_a &= 0\;.\nonumber
\label{C.12}
\end{align}
We then define the $(n-1)\,$-forms $T_{a\alpha}$ 
and the $n\,$-forms $U_{a\alpha}$ via
\begin{align}
 T_{a\alpha}&:=\star A^*_a\,\Theta_\alpha+k^{n-2}_a[\Theta_\alpha]^1\;,
 \\
 U_{a\alpha}&:= d^nx C^*_a\Theta_\alpha+\star
A^*_a[\Theta_\alpha]^1+k^{n-2}_a[\Theta_\alpha]^2\,.
\label{C.14}
\end{align}
They satisfy 
\begin{align}
 s U_{a\alpha} + dT_{a\alpha} & = (-)^{n} k^{n-2}_a \,N^2_\alpha\;,\nonumber
 \\
s T_{a\alpha} + d (k^{n-2}_a\Theta_\alpha) & =0\;,
\label{C.16}\\
s (k^{n-2}_a\Theta_\alpha) & = 0\,.\nonumber
\end{align}

\item The characteristic classes\footnote{By an abuse of terminology,
    in the present context, characteristic classes are particular
    elements of $\cal B\,$ given by $\mathfrak{g}\,$-invariant
    polynomials in the curvature two-forms $F^I$.} in form degree
  $n-1\,$ that are on-shell trivial in $\cI\,$ can only exist for $n$
  odd since the curvatures are two-forms. Let
  $P^{n-1}_{\bar {\cal A}}(F)\approx dI^{n-2}_{\bar\cA}\,$,
  $I^{n-2}_{\bar {\cal A}}\in \cI\,$, denote a basis of such
  characteristic classes: $P^{n-1}(F)\approx dI^{n-2}$,
  $I^{n-2}\in \cI$ implies that
  $P^{n-1}=\lambda^{\bar\cA} P_{\bar\cA}$, with
  $\lambda^{\bar\cA} P_{\bar\cA}=0 \Rightarrow \lambda^{\bar\cA}=0\,$.
  On the other hand, as for any characteristic class, 
  \begin{align}
    P^{n-1}_{\bar \cA} = dq_{\bar \cA}^{n-2}\,
    \label{Pisdexact}
  \end{align}
  for an $(n-2)\,$-form $q_{\bar \cA}^{n-2}\not\in \cal I\,$, of the
  ``Chern-Simons'' type.  Together with the defining property that
  $P^{n-1}_{\bar \cA}\approx dI^{n-2}_{\bar\cA}\,$, one deduces that
  $q_{\bar \cA}^{n-2}-I^{n-2}_{\bar \cA}$ is a non-covariantizable
  element of $H^{-2,n}(s|d)\,$:
  $q_{\bar \cA}^{n-2}-I^{n-2}_{\bar \cA}=\lambda^\cA_{\bar\cA}k_\cA$
  $=: k_{\bar \cA} \,$, see subsection \ref{sec:comm-char-cohom}.
  When taking into account \eqref{noncovariantcharactn-2a} and
  defining
  $\star A^*_{\bar\cA}:=\lambda^\cA_{\bar\cA}\star A^*_{\cA}\,$,
  $d^nx C^*_{\bar\cA}:=\lambda^\cA_{\bar\cA}\,d^nx C^*_{\cA}\,$, one
  gets the following descent equations
 \begin{align}
     sd^nx C^*_{\bar\cA} + d\star A^*_{\bar \cA} & = 0\;,\nonumber
     \\
     s\star A^*_{\bar \cA}+d(q_{\bar\cA}^{n-2}-I_{\bar\cA}^{n-2})&=0
     \;,\label{C.20}\\
     sq^{n-2}_{\bar\cA}+dr^{n-3}_{\bar\cA} & = 0\;.\nonumber
 \end{align}
 Note that every $P^{n-1}_{\bar \cA}$ gives rise to
 non-covariantizable characteristic cohomology in degree $n-2$, and is
 thus related to $d^nx C^*_\cA$, but at this stage there could be more
 of the latter. A consequence of the theorem will be that there is not
 and that the indices $\cA$ and $\bar \cA$ can be identified. This is
 analyzed in section \ref{sec:cons-char-cohom}.

\item Let $\{N_\gamma\}$ be a basis for linear combinations of the
  elements of the form $P^{n-1}_{\bar\cA}(F)\Theta_\alpha\in H(s)$
  that are at the same time obstructions to lifts in form degree $n-1$
  in the small algebra. A basis of the latter is denoted by $\{N_i\}$, 
  see \cite{Barnich:1994db}, section 10.5. 
  One can choose the basis such that the
  $N_\gamma$'s contain in particular the characteristic classes
  $P^{n-1}_{\bar\cA}(F)$ themselves, in which case
  $\Theta_\alpha = 1\,$.  Indeed, every characteristic class is an
  obstruction to the lift of an element in $\cal B\,$, which amounts
  to saying that $\{N_i\}$ contains a basis of all $P(F)$'s.  By the
  definition of $\{N_\gamma\}\,$, one has
\begin{align}
    N_\gamma & := k^{\bar\cA\alpha}_\gamma
  P^{n-1}_{\bar\cA}\Theta_\alpha =k^{\bar\cA\alpha}_\gamma(dI^{n-2}_{\bar\cA}-s\star
  A^*_{\bar \cA})\Theta_\alpha = 
  \nonumber \\
  &  \qquad = 
  k^{\bar\cA\alpha}_\gamma[d(I^{n-2}_{\bar\cA}\Theta_\alpha)-s(\star
  A^*_{\bar \cA}\Theta_\alpha - I^{n-2}_{\bar\cA}[\Theta_\alpha]^1)]\,, 
  \label{C.22}
  \end{align}
  where the second equality requires \eqref{Pisdexact} and 
  \eqref{C.20} and the third one uses the invariance of the
  basis $\{\Theta_\alpha\}$ together with \eqref{C.8}.  
  
  On the other hand, because $N_\gamma \in\{N_i\}$ one has
  $k^{\bar\cA\alpha}_\gamma
  P^{n-1}_{\bar\cA}\Theta_\alpha\equiv N_\gamma = 
  s b^{n-1}_{\gamma}+dB^{n-2}_\gamma\,$.  There then exist
  $b^n_\gamma$ such that 
  \begin{align}
      k^{\bar {\cal A} \alpha}_\gamma \, 
  P^{n-1}_{\bar {\cal A}} [\Theta_\alpha]^1 & = db^{n-1}_\gamma + s b^n_\gamma\;.
  \end{align}
  This follows directly from the analysis of the descent equations in
  the small algebra of \cite{Barnich:1994db}, subsection 10.6. One
  has
  $(s+d)M_{r_1\dots r_K|s_1\dots s_N}(q(C+A,F),f(F)) = \sum_r f_r
  \frac{\partial}{\partial q_r}M_{r_1\dots r_K|s_1\dots
    s_N}(q(C+A,F),f(F))\,$.  The equation for $N_\gamma$ is the one in
  form degree $\underbar{s}+2m(r_1)\,$, while the searched for
  relation is the one in form degree $\underbar{s}+2m(r_1)+1\,$.

Now, defining 
\begin{align}
   W_\gamma & := b^{n-1}_\gamma+k_\gamma^{\bar\cA\alpha}(-
  I^{n-2}_{\bar\cA}[\Theta_\alpha]^1+\star A^*_{\bar
    \cA}\Theta_\alpha)\;,
    \label{C.24}
    \\
   R_{\gamma}& :=b^n_\gamma + k^{\bar\cA\alpha}_\gamma(-I^{n-2}_{\bar\cA}[\Theta_\alpha]^2+
   \star A^*_{\bar \cA}[\Theta_\alpha]^1 + 
   \star C^*_{\bar \cA}\Theta_\alpha)\;, 
   \label{C.25}
\end{align}
 it is direct to check that the following equations are satisfied:
\begin{align}
  s R_{\gamma} + d W_\gamma &= 
  k^{\bar\cA\alpha}_\gamma I^{n-2}_{\bar\cA} N^2_\alpha\;,
   \label{C.26}\\
    s W_\gamma + d (B^{n-2}_\gamma 
  - k^{\bar\cA\alpha}_\gamma I^{n-2}_{\bar \cA}\Theta_\alpha) & = 0\;.
  \label{C.27}
\end{align}

\item We can repeat the analysis of the previous item in form degree
  $n$ and consider characteristic classes $P^n(F)$ that are on-shell
  $d\,$-trivial in $\cI\,$. They can only exist for even $n$.  We
  denote a basis of such classes by $P_A(F)$:
  $P_A(F)\approx dI^{n-1}_A$, $I^{n-1}_A\in \cI\,$ and
  $P^n(F)\approx dI^{n-1}\,$, $I^{n-1}\in \cI$
  $\Rightarrow P^n=\lambda^AP_A\,$, with
  $\lambda^A P_A=0 \Rightarrow \lambda^A=0\,$.  As for any
  characteristic class, $P_A = dq_A^{n-1}$. This allows us to
  construct the non-covariantizable on-shell conserved currents
  \begin{equation}
j_A:=q_A^{n-1}-I^{n-1}_A\,\label{eq:60}
\end{equation}
with associated $K_A$ linear in the antifields such that
$sK_A + dj_A = 0\,$.
Let $N_\Gamma$ denote a basis for the elements of the form
$P_A(F)\Theta_\alpha\in H(s)$ that are at the same time obstructions
to lifts in form degree $n$ in the small algebra. We have
$N_\Gamma=k_\Gamma^{A\alpha}(dI^{n-1}_A - sK_A)\Theta_\alpha
=k^{A\alpha}_\Gamma[d(I^{n-1}_A\Theta_\alpha) - s(K_A\Theta_\alpha -
I^{n-1}_A[\Theta_\alpha]^1)]\,$. At the same time,
$N_\Gamma = s b^n_{\Gamma} + dB^{n-1}_\Gamma\,$.  We then obtain
  \begin{equation}
    s W_\Gamma + d(B^{n-1}_\Gamma-k^{A\alpha}_\Gamma
   I^{n-1}_A\Theta_\alpha) = 0\;, \quad
      W_\Gamma  := b^n_\Gamma - k_\Gamma^{A\alpha}\,
     ( I^{n-1}_A[\Theta_\alpha]^1 - K_A\Theta_\alpha)\;.
     \label{C.29}
  \end{equation}

\item A basis for gauge-invariant, non-trivial Noether currents
is given by elements 
\begin{align}
  j_\Delta=d^{n-1}x_\mu\, j^\mu_\Delta ~ \in \cI\,.  
\end{align}
Each of these elements satisfies $d j_\Delta\approx 0$ and
$dI^{n-1}\approx 0 \,\Rightarrow I^{n-1}\approx \lambda^\Delta
j_\Delta+d\omega^{n-2}\,$ with
$\lambda^\Delta j_\Delta\approx d\eta^{n-2}\,\Rightarrow
\lambda^\Delta=0$. For every $\Delta$ there exists an element
$K_\Delta$ at antifield number one such that
$d j_\Delta + sK_\Delta=0\,$. Defining
$V_{\Delta\alpha} :=
K_\Delta\Theta_\alpha+j_\Delta[\Theta_\alpha]^1\,$, the following
descent equations hold,
\begin{equation}
s V_{\Delta\alpha}+d(j_\Delta\Theta_\alpha)=0\,,\quad
s(j_\Delta\Theta_\alpha)=0\,. \label{eq:10}
\end{equation}
\end{enumerate}

\subsection{Statement of the theorem}
\label{sec:theorem-11.1-bis}

The theorem generalizes theorem 11.1 of \cite{Barnich:2000zw}. It
decomposes into four items. The first characterizes the general
solution of the cocycle condition $s\omega^p + d\omega^{p-1}= 0$, up
to trivial ones. The second makes precise their linear
dependence. The third item discusses the decomposition of an invariant
form that is $d\,$-exact on-shell, while the fourth item parametrizes
characteristic classes that are trivial in
$H_{\rm char}^*(d,\cal I)\,$. Equivalence in $H^{g,n}(s|d)$ is denoted
by $\sim$: $\omega^{\prime g,p}\sim \omega^{g,p}$ if $\omega^{\prime
  g,p}= \omega^{g,p}+s\eta^{g-1,p}+d\eta^{g,p-1}$, for some
$\eta^{g-1,p},\eta^{g,p-1}$. 
%\newpage

\noindent \textbf{\emph{Theorem}}
\begin{enumerate}
\item[(i)] The general solution of the cocycle condition 
$s\omega^p + d\omega^{p-1}= 0 $ is given by
\begin{multline}
 \omega^p \sim I^{p\alpha}\Theta_\alpha + B^p
+\delta^p_{n-1}( \lambda^{a\alpha}T_{a\alpha}
+\lambda^\gamma W_\gamma) 
  \label{eq:1} \\
 +\delta^p_n \left(I^{n-1\alpha}_{(\mu)}[\Theta_\alpha]^1+\hat
  B^{n}_{(\mu)}+b^n_{(\mu)} +\mu^{a\alpha}U_{a\alpha}+\mu^\gamma
  R_\gamma+K^{\alpha}_{(\mu)}\Theta_\alpha +\lambda^{\Delta\alpha}
  V_{\Delta\alpha}+\lambda^\Gamma W_\Gamma\right)\,,
\end{multline}
where $\mu^{a\alpha}$, $\mu^\gamma$ is the most general solution to
the obstruction equation, 
\begin{equation}
[\mu^{a\alpha}(-)^{n}k^{n-2}_a
+\mu^{\gamma}k_\gamma^{\bar\cA\alpha}I^{n-2}_{\bar\cA}]N^{2}_\alpha+N^n_{(\mu)}+
dI^{n-1\alpha}_{(\mu)}\Theta_\alpha+s(K^{\alpha}_{(\mu)}\Theta_\alpha)=0,\label{eq:5}
\end{equation}
and where elements with the subscript $(\mu)$ vanish when the $\mu$'s vanish.

\item[(ii)] 
A solution as in \eqref{eq:1} is trivial in $H^{g,p}(s|d)$,
$\omega^p=s\eta^{p}+d\eta^{p-1}$, or more explicitly,
\begin{multline}
  \label{eq:2}
0 \sim I^{p\alpha}\Theta_\alpha + B^p
+\delta^p_{n-1}( \lambda^{a\alpha}T_{a\alpha}
+\lambda^\gamma W_\gamma) 
\\
 +\delta^p_n \left(I^{n-1\alpha}_{(\mu)}[\Theta_\alpha]^1+\hat
  B^{n}_{(\mu)}+b^n_{(\mu)} +\mu^{a\alpha}U_{a\alpha}+\mu^\gamma
  R_\gamma+K^{\alpha}_{(\mu)}\Theta_\alpha +\lambda^{\Delta\alpha}
  V_{\Delta\alpha}+\lambda^\Gamma W_\Gamma\right)\,,
\end{multline}
if and only if 
\begin{multline}
 0 = \lambda^\Gamma=\lambda^{\Delta\alpha}=\lambda^{a\alpha}
=\lambda^\gamma  =\mu^{a\alpha}
=\mu^{\gamma}=K_{(\mu)}=b^n_{(\mu)}=\hat
       B^n_{(\mu)}=I^{n-1\alpha}_{(\mu)}=B^p\;,
  \\ 
I^{p\alpha}\Theta_\alpha  \approx N^p+dI^{p-1\alpha}\Theta_\alpha
   +\delta^p_n [\rho^{a\beta}(-)^{n}k^{n-2}_a
   +\rho^{\gamma}k_\gamma^{\bar\cA\beta}I^{n-2}_{\bar\cA}]N^2_\beta\;.
   \label{eq:2biss}
\end{multline}

\item[(iii)] If $I^p\in \cal I$ ($p>0$) is trivial in 
$H^p_{\rm char}(d,\Omega)\,$, then it is the sum of a characteristic
class and a piece which is trivial in $H^p_{\rm char}(d,\cal I)\,$: 
\begin{equation}
   p>0~:~ \quad I^p\approx d\eta^{p-1}~\iff ~
   I^p \approx P^p(F) + dI^{p-1}\; ;
\label{eq:3}   
\end{equation}

\item[(iv)] No non-trivial characteristic class in form degree
  strictly less than $n-1$ is trivial in $H_{\rm char}(d,\cI)$, 
\begin{equation}
    \label{eq:4}
   p<n-1:\quad P^p(F)\approx dI^{p-1}\Longrightarrow P^p(F)=0. 
  \end{equation}
  Note that, by definition, in form degrees $n-1$ and $n$,
  characteristic classes that are trivial in $H_{\rm char}(d,\cI)$ can
  be written as linear combinations of the bases elements
  $P_{\bar\cA},P_A\,$:
  \begin{equation}
P^{n-1}(F)\approx dI^{n-2}\Longrightarrow P^{n-1}=\lambda^{\bar\cA}
  P_{\bar\cA}\,,\quad 
  P^{n}(F)\approx dI^{n-1}\Longrightarrow P^{n}=\lambda^A
  P_A\,\label{eq:59}.
\end{equation}
  
\end{enumerate}

\subsection{Structure of characteristic cohomology in degree
    n-2}
\label{sec:cons-char-cohom}

We have shown that every characteristic class that is weakly $d$ exact
in ${\cal I}$ for form degrees $n-1$ and $n$ gives rise to
non-covariantizable characteristic cohomology in degrees $n-2$ and
$n-1\,$. What remains to be analyzed is the converse.  Characteristic
cohomology in form degree $n-2$ is isomorphic to $H^{-2,n}(s|d)$ (see
e.g.~Theorem 7.1 of \cite{Barnich:2000zw}; there can be no constant
since we assume $n>2\,$). Only $U_{a\alpha}$ and $R_\gamma$ contain
ghost number $-2$ pieces, see \eqref{C.14} and \eqref{C.25}, and this
requires that the pure-ghost part of them should vanish, i.e.,
$\Theta_\alpha=1\,$.  In this case $N^2_\beta=0$ and equation
\eqref{eq:5} imposes no restrictions on $\mu^{a},\mu^\gamma\,$.  It
follows that
$\omega^{-2,n}\sim \mu^a d^nx C^*_a+\mu^\gamma k^{\bar\cA}_\gamma d^nx
C^*_{\bar\cA}\,$.  Since we can choose the basis $N_\gamma$ to include
the $P_{\bar\cA}\,$ and since, with $\Theta_\alpha=1\,$, the
characteristic classes $P_{\bar {\cal A}}$'s are actually equivalent
to the $N_\gamma$'s from the very definition of the latter, we can
take $k^{\bar\cA}_\gamma=\delta^{\bar\cA}_\gamma$ and therefore
    \begin{align}
    \omega^{-2,n}\sim \mu^a d^nx C^*_a+\mu^{\bar\cA} d^nx 
    C^*_{\bar\cA}\;.
    \end{align}
    Since the first piece corresponds to covariantizable
    characteristic cohomology, we have shown that non-covariantizable
    characteristic cohomology in degree $n-2$ is exhausted by the
    classes related to $d^nx C^*_{\bar\cA}$ which are explicitly given
    by $B_\gamma^{n-2}-\delta^{\bar\cA}_\gamma I^{n-2}_{\bar\cA}\,$,
    see equation \eqref{C.20}. The former term satisfies
    $P_{\bar\cA}=dB^{n-2}_{\bar\cA}$ where
    $B^{n-2}_{\bar\cA}\equiv q^{n-2}_{\bar\cA}\,$.  Indeed, since
    $\Theta_\alpha=1\,$,
    $k^{\bar{\cal A}}_\gamma P_{\bar{\cal A}}= N_\gamma =
    dB^{n-2}_\gamma + s b^{n-1}_\gamma$ is satisfied with
    $b^{n-1}_\gamma=0\,$ and we recall that we chose
    $k^{\bar{\cal A}}_\gamma =\delta^{\bar{\cal A}}_\gamma\,$.  We
    have thus shown that non-covariantizable characteristic cohomology
    is exhausted by characteristic classes that become trivial
    on-shell and that, by a change of basis, the indices $\bar\cA$ and
    $\cA$ can be identified.

    In form degree $n-2$, we thus have a direct sum decomposition of
    characteristic cohomology into $U$-type, which is covariantizable
    and related to $\mu^a$ with associated elements of $H^{-2,n}(s|d)$
    of depth $2$, and $R$-type, which is non-covariantizable and
    related to $\mu^{\bar{\cal A}}$, with associated elements of
    $H^{-2,n}(s|d)$ of depth at least $3$ that can only exist in odd
    spacetime dimensions.

\subsection{Structure of characteristic cohomology in degree
  n-1}\label{sec:cons-char-cohombis}

Characteristic cohomology in form degree $n-1$ is isomorphic to
$H^{-1,n}(s|d)\,$. From expression \eqref{eq:1} one can see that the
elements of $\omega^n$ with antifield number 1 but not $2$ should be
multiplied by $\Theta_\alpha=1\,$. In this case, it follows from the
definition of $N_\Gamma$ that the $P_A$'s are equivalent to the
$N_\Gamma$'s and we can choose $k^A_\Gamma=\delta^A_\Gamma$. Hence,
   \begin{align}
    \omega^{-1,n}\sim \lambda^\Delta K_\Delta+\lambda^A
       K_A+[\mu^{a\alpha}U_{a\alpha} +\mu^\gamma
       R_\gamma+K_\mu^{n\alpha}\Theta_\alpha]^{-1}  \;. 
   \end{align}
   This leads to the following decomposition:
   
   $V$-type corresponds to the elements $\lambda^\Delta K_\Delta$ of
   $H^{-1,n}(s|d)$ that have depth $1$ and the associated
   covariantizable 
   characteristic cohomology is $\lambda^\Delta j_\Delta$. 

   $W$-type corresponds to the elements $\lambda^A K_A$ of
   $H^{-1,n}(s|d)$ that have depth at least $2$ with associated
   non-covariantizable characteristic cohomology given by
   $\lambda^A (B^{n-1}_A-I_A^{n-1})\,$. Here $B^{n-1}_A:=q^{n-1}_A$
   coincides with $ B^{n-1}_\Gamma \equiv k^A_\Gamma q^{n-1}_A\,$, see
   item 5 in section \ref{sec:ingredients}.

   In order to work out the last terms, the ghost number $-1$ part of
   the $\mu$ part, we need to combine the pieces in $U_{a\alpha}$
   having strictly positive antifield number (1 and 2) with
   corresponding $\Theta_\alpha$'s given by an abelian ghost
   $C^{I_{ab}}$, so that $[\Theta_\alpha]^2=0$ and
   $N^2_\alpha \equiv F^{I_{ab}}\,$. For terms related to
   $\mu^\gamma\,$ to exist, one needs to be in odd spacetime
   dimensions so that the
   $N_\gamma := k^{\bar\cA\alpha}_\gamma P_{\bar\cA}\Theta_\alpha$ can
   exist. One should also have that
   $k^{\bar\cA\alpha}_\gamma \Theta_\alpha$ be in ghost number $1\,$
   --- see the expression for $R_\gamma\,$ in \eqref{C.25}.  On the
   one hand, from eq. (10.18) of \cite{Barnich:1994db} and eq. \eqref{B.38b},
   \begin{equation*}
   k^i N_i = F^{K_{ab}}\d_{K_{ab}}\left[ P^{n-3}(F)\half 
   k_{[I_{ab}J_{ab}]}C^{I_{ab}}C^{J_{ab}}\right]
   =P^{n-3}(F)F^{I_{ab}}k_{[I_{ab}J_{ab}]}C^{J_{ab}}\;.
   \end{equation*}
   On the other hand, this needs to be equal to a linear combination
   of $P_{\bar\cA}C^{I_{ab}}\,$.  The gauge group thus needs to
   contain at least two abelian factors and hence two different
   $P_{\bar\cA}(F)$'s containing abelian field strengths. In turn,
   this requires at least two different free abelian factors ---
   recall that the $P_{\bar\cA}(F)$'s are related to the
   characteristic cohomology in degree $n-2$ and hence to
   $C^*_{\cal A}\,$. Putting everything together, the expression for
   $[\mu^{a\alpha}U_{a\alpha}+\mu^\gamma
   R_\gamma+K^{n\alpha}_\mu\Theta^\alpha]^{-1}$ looks as
   follows
   \begin{align}
    \mu^a_{I_{ab}}(d^nx C^*_a
   C^{I_{ab}}+\star A^*_a A^{I_{ab}})+\mu^\gamma 
    k^{\bar\cA}_{\gamma I_{ab}}
    (d^nx C^*_{\bar\cA}C^{I_{ab}}+\star A^*_{\bar\cA}A^{I_{ab}})+K^n_\mu   \;.
   \end{align}

   The condition that this defines additional $H^{-1,n}(s|d)$
   cohomology and thus additional characteristic cohomology in form
   degree $n-1$ is the existence of particular $P^n_\mu$,
   $I^{n-1}_\mu$, $K^n_\mu$ allowing one to solve the obstruction
   equation,
   \begin{equation}
   [(-)^{n}\mu^a_{I_{ab}}k_a^{n-2}+\mu^\gamma k^{\bar\cA}_{\gamma
   I_{ab}}I^{n-2}_{\bar\cA}]F^{I_{ab}}+P^n_\mu +dI^{n-1}_\mu+s
   K^n_\mu=0\label{eq:6}. 
 \end{equation}
The associated characteristic cohomology in degree $n-1$ is determined
by the antifield independent part of
\begin{align}
 \mu^{a}_{I_{ab}}(k_a^{n-2}A^{I_{ab}}+\star A^*_a
C^{I_{ab}})+\mu^\gamma [b_\gamma^{n-1}+k^{\bar\cA}_{\gamma
  I_{ab}}(-I^{n-2}_{\bar\cA}A^{I_{ab}}+\star
A^*_{\bar\cA}C^{I_{ab}})]\;,   
\end{align}
depends explicitly on $A^{I_{ab}}$. The argument that no allowed
redefinition makes these additional characteristic cohomology classes
invariant goes as follows: suppose that one of these classes could be
made equivalent to a combination of $j_\Delta$. Then the BRST
extension $\omega^{0,n-1}$ that contains that class would satisfy
$\omega^{0,n-1}\sim \lambda^\Delta j_\Delta$. In turn, this implies
that the associated $\omega^{-1,n}\sim \lambda^\Delta K_\Delta$. But
part (ii) of the theorem then shows that this implies that the
coefficients of all these terms vanish separately.

$U$-type corresponds to solutions with vanishing $\mu^\gamma$ but
non-vanishing $\mu^{a}_{I_{ab}}$. They are thus related to
covariantizable characteristic cohomology in degree $n-2$ through the
parameters $\mu^{a}_{I_{ab}}$. They have depth $2$. Note that such
solutions may involve $K_\mu$'s that vanish when the
$\mu^{a}_{I_{ab}}$ do.

$R$-type corresponds to solutions with non-vanishing
$\mu^\gamma$'s. They are related to non-covariantizable characteristic
cohomology in degree $n-2$ through the parameters
$\mu^\gamma k^{\bar A}_{\gamma I_{ab}}$. They have depth at least
$3$. Such solutions may involve $K_\mu$'s and $\mu^a_{I_{ab}}(d^nx C^*_a
   C^{I_{ab}}+\star A^*_a A^{I_{ab}})$'s that vanish when the
   $\mu^\gamma$ vanish. 

   In particular, for pure abelian Yang-Mills theory there is no
   $R$-type, so that the additional non-covariantizable characteristic
   cohomology is of $U$-type and reduces to
   $\mu^a_{I_{ab}} k^{n-2}_aA^{I_{ab}}$. This class then contains the
   rigid symmetries corresponding to the rotation of free abelian
   vector fields among themselves, see section
   \ref{sec:pure-abelian-yang} and \cite{Barnich:2017nty} for further
   discussions.

\subsection{Structure of infinitesimal deformations}
\label{sec:interpr-results}

Infinitesimal deformations are described by $H^{0,n}(s|d)$ and are
thus captured, cf. Item (i) of the theorem, by the nontrivial cocycles
of $s$ modulo $d\,$, at ghost number zero and top form degree $n\,$:
\begin{align}
\omega^{0,n}\sim I^{n} +B^{0,n}+V^{0,n}+W^{0,n}
+[\mu^{a\alpha}U_{a\alpha}+\mu^\gamma
  R_\gamma+K^{n\alpha}_\mu\Theta_\alpha+I^{n-1\alpha}_\mu[\Theta_\alpha]^1+\hat
  B^{n}_\mu+b^n_\mu ]^0\;,
  \nonumber
\end{align}
where the ghost number is indicated in the superscript of the last term 
between square bracket; the form degree of the quantity 
between square bracket is $n\,$. 
According to item (ii) of the theorem, such a solution is 
non-trivial whenever $I^n\napprox
d\omega^{n-1}$, which is equivalent to saying that $I^n$
is not expressed as in \eqref{eq:2biss},
and the other terms in the expression of $\omega^{0,n}$ do not
vanish.  
 
More precisely, non-trivial solutions $\omega^{0,n}$ can be decomposed 
into the following linearly independent types:

\begin{itemize}

\item $I$-type: non trivial elements $I^n= d^nx\
  I_{inv}([F,\psi])\in\mathcal I\,$;

\item $B$-type: $B^{0,n}$ is a linear combination of the independent
  Chern-Simons $n$ forms 
  \begin{align}
    [M_{r|s_1\ldots s_N}]^{\underline{s}+2m(r)-1} 
    & = [\theta_r]^{2m(r)-1}\,f_{s_1}\ldots f_{s_N}\;,  
  \end{align}
  therefore they only arise in odd spacetime dimensions $n=\underline{s}+2m(r)-1\,$;

\item $V$-type: $V^{0,n}=\lambda^{\Delta}_{I_{ab}}(K_\Delta C^{I_{ab}}+j_{\Delta}
  A^{I_{ab}})$, they correspond to standard gaugings obtained from
  minimal coupling of abelian gauge fields to covariantizable
  conserved Noether currents;

\item $W$-type: for $W^{0,n}$, the spacetime dimension needs to be
  even and one needs a relation
  $k^\Gamma N_\Gamma=k^i N^n_i=k^{A\alpha}P^n_A(F)\Theta_\alpha$ at
  ghost number $1\,$, so that $b^n_\Gamma$ in the expression of
  $W_\Gamma$ in \eqref{C.29} has ghost number zero.  The left hand
  side of the previous equation is of the form
  $F^{K_{ab}}\d_{K_{ab}}\left( P^{n-2\,\Sigma}(F)\half k_{\Sigma
      \,I_{ab}J_{ab}}C^{I_{ab}}C^{J_{ab}}\right)=P^{n-2\,\Sigma}(F)
  F^{I_{ab}}k_{\Sigma\,I_{ab}J_{ab}}C^{J_{ab}}\,$. This needs to be
  equal to a linear combination of $P_{A}(F)C^{I_{ab}}\,$. The gauge
  group thus needs to contain at least two abelian factors and two
  different $P_{A}(F)$'s containing abelian field strengths.  For an
  example see equation (12.4) of \cite{Barnich:2000zw}.  Although it
  illustrates these classes, the Lagrangian used there is not covered
  by theorem 11.1 of \cite{Barnich:2000zw} since it contains free
  abelian gauge fields. It fits perfectly however in the present
  context, and is treated in more detail in \cite{Barnich:2017nty}.
  
\item $U$-type correspond to solutions to the obstruction equation
  \eqref{eq:6} with vanishing $\mu^\gamma$ and non-vanishing
  $\mu^{a\alpha}$. In ghost number $0$, they are related to
  covariantizable characteristic cohomology in degree $n-2$ through
  the parameters $\mu^a_{I_{ab}J_{ab}}$ and contain
  \begin{align}
  [\mu^{a\alpha}U_{a\alpha}]^0=(d^nx C^*_a +\star
  A^*_aA^{K_{ab}}\d_{K_{ab}}+ \half k_a^{n-2}
  A^{K_{ab}}A^{L_{ab}}\d_{L_{ab}}\d_{K_{ab}})
  \tfrac{1}{2}\,\mu^{a}_{[I_{ab}J_{ab}]}C^{I_{ab}}C^{J_{ab}}\;,
  \nonumber
  \end{align}
  where $\d_{I_{ab}}=\partial/\partial C^{I_{ab}}$.
  In general, these terms might need to be supplemented by terms of the form
  $K^n_{\mu\, I_{ab}}C^{I_{ab}}+I^{n-1}_{\mu\, I_{ab}}A^{I_{ab}}+\hat
  B^n_\mu+b^n_\mu$ that vanish when the $\mu^{a}_{[I_{ab}J_{ab}]}$
  vanish.  

\item $R$-type correspond to solutions to the obstruction equation
  \eqref{eq:6} with non-vanishing $\mu^\gamma$'s. The dimension $n$ of
  spacetime needs to be odd for the existence of $N^{n-1}_\gamma$ and
  we must have that $b^n_\gamma$ has ghost number zero, which implies
  that $gh(b^{n-1}_\gamma)=1$ and hence $gh(N_\gamma)=2\,$. Since the
  $\theta_r$'s have odd ghost number, the ghost degree of $N_\gamma$
  is carried by the product of two abelian ghosts. Therefore, from eq. (10.18)
  of \cite{Barnich:2000zw} and from \eqref{B.38b}, the linear combination
  $k^\gamma N_\gamma = k^iN_i\,$ should have the form
  \begin{align}
    k^iN_i = F^{I_{ab}}\d_{I_{ab}}\left( P^{n-2\,\Sigma}(F)\tfrac{1}{6}\,
  k_{\Sigma \,I_{ab}J_{ab}K_{ab}}C^{I_{ab}}C^{J_{ab}}C^{K_{ab}}\right)
  &  
  \nonumber \\
  ~=~ \tfrac{1}{2}\,P^{n-2\,\Sigma}(F)&
  F^{I_{ab}}k_{\Sigma\,I_{ab}J_{ab}K_{ab}}C^{J_{ab}}C^{K_{ab}}\,,  
  \nonumber
  \end{align}
  and this needs to be equal to a linear combination of 
  $P_{\bar {\cal A}}(F)C^{I_{ab}}C^{J_{ab}}\,$. 
  The gauge group thus needs to contain at least three abelian factors 
  and three different
  $P_{\bar {\cal A}}(F)$'s containing abelian field strengths.
  In turn, this requires at least 3 free abelian factors.
  
  Finally, we note from \eqref{C.25} that $[\mu^\gamma R_\gamma]^{0}\,$ terms contains a piece linear in
  the antighost $C^*_{\bar{\cal A}}\,$, and that these terms might need to
  be supplemented by terms of the form $[\mu^{a\alpha}U_{a\alpha}]^0$
  and  $K^n_{\mu\, I_{ab}}C^{I_{ab}}+I^{n-1}_{\mu\, I_{ab}}A^{I_{ab}}+\hat
  B^n_\mu+b^n_\mu$ that vanish when the $\mu^\gamma$
  vanish.
\end{itemize}
A similar analysis can be performed for anomaly candidates by spelling
out the classes in ghost number $1$. 

\section{Applications}
\label{sec:appl-known-models}

\subsection{No free abelian factors}
\label{sec:no-free-abelian}

If the gauge group contains no free abelian factors, there is no
characteristic cohomology in degree $n-2$,
$H^{n-2}_{\rm char}(d)\simeq H^{-2,n}(s|d)$ vanishes. This implies the
absence of $k_a^{n-2}$ and the associated $T_{a\alpha}$ and
$U_{a\alpha}$ on the one hand, and also the absence of $P_{\bar\cA}$,
and the associated $W_\gamma$ and $R_\gamma$, on the other. In other
words,
$\lambda^{a\alpha}=\lambda^\gamma=\mu^{a\alpha}=\mu^\gamma=
I^{n-1\alpha}_\mu=\hat B^n_\mu=b^n_\mu=0=\rho^{a\beta}=\rho^\gamma$ in
the theorem, which therefore reduces to theorem 11.1 of
\cite{Barnich:2000zw}.

\subsection{Chern-Simons theory}
\label{sec:chern-simons-theory}

We consider pure Chern-Simons theory, as was done in the section 14 of
\cite{Barnich:2000zw}, but now as a particular case of the present
approach, instead of an AKSZ-type approach with complete ladder fields
\cite{Carvalho:1995uu,Carvalho:1995ut,Barnich:2009jy}. The action is
given by
\begin{equation}
\label{eq:7}
S_{CS}=\int g_{IJ} (\half A^I dA^J+\frac{1}{6} f^I_{KL} A^JA^KA^L).
\end{equation}
The first question is about characteristic cohomology in degree $n-2$:
it can be represented by $d^nx C^*_{I_{ab}}$ and is associated with
each (free\footnote{In pure Chern-Simons theory, there is no
  distinction between abelian and free abelian factors.})  abelian
factor. Furthermore, as can be seen from
$s A^{*\mu}_I=\half g_{IJ}\epsilon^{\mu\nu\rho}{\cal
  F}_{\nu\rho}^J+C^Jf^K_{JI}A^{*\mu}_K\,$, every characteristic class
in form degree $2=3-1$ associated with an abelian factor is trivial
on-shell:
$F^{I_{ab}}=\half F^{I_{ab}}_{\mu\nu} dx^\mu dx^\nu=-g^{I_{ab}J_{ab}}s
\star A^{*}_{J_{ab}}=-s\star A^{*I_{ab}}\,$.  It follows that all
abelian indices are $\bar\cA$ indices.  There is thus no $T_{a\alpha}$
nor $U_{a\alpha}\,$:
$\lambda^{a\alpha}=0=\mu^{a\alpha}=\rho^{a\beta}\,$.  We also see that
$I^{n-2}_{\bar\cA}=0\,$.  The $M_{r_1\dots r_K|s_1\dots s_N}$ of
equation (10.16) of \cite{Barnich:2000zw}, the $M$'s in \eqref{B.38b} 
and the associated $N_{r_1\dots r_Ks_1\dots
  s_N}:=\sum_{r:m(r)=m(r_1)}f_r\,\ddl{M_{r_1\dots r_K|s_1\dots
    s_N}}{\theta_r}$ and $N$'s can be split according
to whether they contain abelian factors or not, $m(r_1)=1$ or
$m(r_1) > 1\,$.  The former constitute the $N_\gamma$'s since they are
given by
$F^{I_{ab}}\d_{I_{ab}}{M_{r_1\dots r_K|s_1\dots s_N}}=-s (\star
A^{*I_{ab}}\d_{I_{ab}}{M_{r_1\dots r_K|s_1\dots s_N}})\,$, cf.~\eqref{C.22}.

Now we need to address the question whether these $N_\gamma$
can be lifted upon adding to $b^{n-1=2}_\gamma$ a contribution outside
of ${\cal B}\,$, thereby obtaining a $W_\gamma\,$.  Condition
\eqref{eq:5} is empty since there is no $\mu^{a\alpha}$ and no
$I^{n-2}_{\bar\cA}$ so all of the $W_\gamma$'s can be lifted to
$R_\gamma\,$ without obstruction, see also \eqref{C.24}--\eqref{C.27}.

This can also be seen differently, as in section 14 of \cite{Barnich:2000zw}:
Introduce $\cC^I=C^I+A^I+\star A^{*I}+d^nx C^{*I}$ so that
$(s+d)\cC^I=f^I_{JK}\cC^K \cC^J\,$ and replace in
$M_{r_1\dots r_K|s_1\dots s_N}\,$ every $\theta_r(C)$ by
$\theta_{r}(\cC)\,$. 
It follows that
$(s+d)M_{r_1\dots r_K|s_1\dots s_N}(\theta(\cC),f)=0\,$.  When
splitting according to the form degree, we have
\begin{align}
M_{r_1\dots r_K|s_1\dots
    s_N}(\theta(\cC),f) & =M_{r_1\dots r_K|s_1\dots
    s_N}(\theta(C),f)+M^{L+1}_{r_1\dots r_K|s_1\dots
    s_N}+M^{L+2}_{r_1\dots r_K|s_1\dots
    s_N}+\dots\;.    
\nonumber
\end{align}
where 
\begin{align}
M^{L+1}_{r_1\dots r_K|s_1\dots
    s_N} & = A^{I_{ab}}\d_{I_{ab}}{M_{r_1\dots r_K|s_1\dots
    s_N}}+\sum_{r: m(r)>1} (A^{I_{nab}}\d_{{I_{nab}}}\theta_r)
    \ddl{M_{r_1\dots r_K|s_1\dots s_N}}{\theta_r}  \;,
 \nonumber \\
M^{L+2}_{r_1\dots r_K|s_1\dots s_N} & 
  = \frac{1}{2}\,
  A^{I_{ab}}A^{J_{ab}}\d_{J_{ab}}\d_{I_{ab}}{M_{r_1\dots r_K|s_1\dots s_N}} 
 \nonumber \\ 
& \quad + \sum_{r: m(r)>1}
  (A^{I_{nab}}\d_{{I_{nab}}}{\theta_r})A^{J_{ab}}\d_{J_{ab}}\ddl{M_{r_1\dots
      r_K|s_1\dots s_N}}{\theta_r} 
 \nonumber \\ 
& \quad +\frac{1}{2}\,\sum_{r: m(r)>1,r':m(r')>1}
  (A^{I_{nab}}\d_{I_{nab}}{\theta_r})(A^{J_{nab}}\d_{J_{nab}}{\theta_{r'}})
  \ddl{M_{r_1\dots r_K|s_1\dots s_N}}{\theta_{r'}\partial \theta_r} 
  \nonumber \\
 & \quad +\sum_{r: m(r)>1} (\star A^{*I_{nab}}\d_{I_{nab}}{\theta_r})
 \ddl{M_{r_1\dots r_K|s_1\dots s_N}}{\theta_r}
 +\star A^{*I_{ab}}\d_{I_{ab}}{M_{r_1\dots r_K|s_1\dots s_N}}\;,    
 \label{D.2}
\end{align} 
with $sM^{L+2}+dM^{L+1}=0\,$ and $\,sM^{L+1}+dM=0\,$. 
As was pointed out in \cite{Barnich:2000zw}, 
the antifield dependent term in $M^{L+2}$ which
contains $\star A^{*I_{nab}}$ can 
be replaced by a term in the small algebra $\cal B\,$. 
We know that $\theta_r(C)$ obeying $s\theta_r(C)=0$ 
can be completed to $q_r(\widetilde{C},f)$ that obeys 
the cocycle relation $\widetilde{s}q_r(\widetilde{C},f)=0\,$
for $\widetilde{s}:=s+d\,$ in three dimensions, since the 
$f_r$'s with $m(r)>1$ are forms of degree higher than 3, 
as recalled in section 10 of \cite{Barnich:2000zw}. 
Therefore, if instead of the $\widetilde{s}\,$-cocycle 
$M_{r_1\dots r_K|s_1\dots s_N}(\theta(\cC),f)$
one uses the $\widetilde{s}\,$-cocycle 
obtained by replacing in 
$M_{r_1\dots r_K|s_1\dots s_N}(\theta,f)$
every $\theta_r$ having $m(r)>1\,$ by the corresponding 
$q_r(\widetilde{C},f)\,$, still keeping the abelian 
$\cC_{I_{ab}}$'s unaffected, one obtains an equivalent 
$\widetilde{s}\,$-cocycle without any non-abelian 
$\star A^*_{I_{nab}}\,$ nor any $\star C^*_{I_{nab}}\,$. 
When this
replacement is made, $W_\gamma$ is given by $M^{L+2}_{r_1\dots
  r_K|s_1\dots s_N}\,$, where all terms correspond to $b_\gamma^{n-1=2}$
  except for the last one in \eqref{D.2}. 
There will be no obstruction to a lift since $M^{L+3}$ exists. 
In other words, $b_\gamma^{n-1=2}$ can be lifted once
more when supplemented by the last term in \eqref{D.2}, 
where the lift of this last term brings in $d^nx C^{*I_{ab}}\,$.
Note that, on account of the form degree, $M^{L+2}$
must correspond to $W^2_\gamma$ and hence $L=0\,$.

Because $n=3$ is odd there are no characteristic classes $P^n_A(F)\,$.
There are also no non-trivial $j_{\Delta}$ since all of them
depend on field strengths and their covariant derivatives, which
vanish weakly. For the same reason $I^{n\alpha}\Theta_\alpha$ reduce
to polynomials in the $\theta_r$ times the volume form $d^3x\,$, 
which is what remains of the $M_{r_1\dots r_K|s_1\dots s_N}$ 
in agreement with \cite{Barnich:2000zw}. 

For ghost number $0$ and form degree $n=3$, we need $M^2$ and in particular 
$b^{n-1=2}_\gamma$ to be of ghost number $1\,$, so that $b^n_\gamma$ 
has ghost number zero and qualifies for an infinitesimal deformation entering 
$R_\gamma\,$. This means that $M$ is of ghost number $3$ and the
only possibility then is $3$ abelian ghosts in order to reproduce 
$N_\gamma = F^{I_{ab}}\d_{I_{ab}}{M_{r_1\dots r_K|s_1\dots s_N}}\,$. 
Therefore one must have
$M=\frac{1}{3!}k_{[I_{ab}J_{ab}K_{ab}]}C^{I_{ab}}C^{J_{ab}}C^{K_{ab}}\,$.
We thus remain with
$S^{(1)}=\int \omega^{0,3}$
with $\omega^{0,3}=B^{0,3}+[\frac{1}{3!}k_{[I_{ab}J_{ab}K_{ab}]}
\cC^{I_{ab}}\cC^{J_{ab}}\cC^{K_{ab}}]^3$
and where $B^{0,3}$ are just the Chern-Simons forms for the non-abelian
factors. The second term is explicitly given by
\begin{equation*}
  [d^nx C^{*I_{ab}}\d_{I_{ab}}+\star
    A^{*I_{ab}}A^{J_{ab}}\d_{J_{ab}}\d_{I_{ab}}
        +\frac{1}{3}A^{I_{ab}}A^{J_{ab}}
A^{K_{ab}}\d_{K_{ab}}\d_{J_{ab}}\d_{I_{ab}}] 
          \frac{1}{3!}k_{I_{ab}J_{ab}K_{ab}}C^{I_{ab}}C^{J_{ab}}C^{K_{ab}}\;.
\end{equation*}
Hence, by construction, $k_{[I_{ab}J_{ab}K_{ab}]}$ is completely
skew-symmetric. When asking for the infinitesimal deformation $S^{(1)}$
to be consistent at second order,
$\half (S^{(1)},S^{(1)})+(S^{(0)},S^{(2)})=0\,$, one finds the Jacobi
identity for
$f^{L_{ab}}_{J_{ab}K_{ab}}=\delta^{L_{ab}I_{ab}}k_{I_{ab}J_{ab}K_{ab}}$
by asking that the term in $C^*_{I_{ab}}$, which is the tail of a non
trivial cohomology class, be absent. There are no further conditions
and there is no need for an $S^{(2)}$.

\subsection{Abelian gauge theories}
\label{sec:abel-gauge-theor}

The gauge group contains only free abelian factors without any
specification of the Lagrangian at this stage, so that abelian
Chern-Simons theory is covered and so is the coset model\footnote{The
  analysis here makes more precise the remark on how to extend the
  analysis for a free abelian Lagrangian of Maxwell type to more
  general Lagrangians at the end of section 13.1 of
  \cite{Barnich:2000zw}.}.  The specificity of the models under
consideration is that the $k^\alpha\Theta_\alpha$ are given by
$\Theta(C^I)\,$, polynomials in undifferentiated free abelian ghosts
$C^I$ --- in this section we drop the subscript $fab$ on the index
$I\,$.  It also follows that $\{C^*_{I}\} = \{ C^*_a,C^*_{\bar\cA}\}$
whose indices are lifted with Kronecker deltas. To define
$W_\gamma\,$, we need a relation
$N=P^\Sigma(F)F^I\d_I\Theta_\Sigma =P_{\bar\cA}\Theta^{\bar\cA}$. In
ghost number $0$, we find
$P^\Sigma(F)F^Ik_{\Sigma I}=k^{\bar\cA}P_{\bar\cA}$, in ghost number
$1$, $P^\Sigma(F)F^Ik_{\Sigma[IJ]}=k^{\bar\cA}_JP_{\bar\cA}$ ... with
$N_\gamma$ the associated basis. This basis has been explicitly given
in \eqref{B.38}. 

If
$W_\gamma=b_{\gamma}^{n-1}+(\star
A^{*\bar\cA}\Theta_{_\gamma\bar\cA}-I^{n-1\bar\cA}A^I\d_I\Theta_{\gamma\bar\cA})$,
we have
\begin{equation*}
\omega^{n-1}\sim I^{n-1\alpha}\Theta_\alpha+B^{n-1}+\star
A^{*a}\Theta_a+k^{n-2a}A^I\d_I\Theta_a+\lambda^\gamma W_\gamma
\end{equation*}
and the obstruction equation \eqref{eq:5} becomes 
\begin{equation}
  \label{eq:8}
  (-)^{n}k^{n-2 a}F^I\d_I\Theta_a+\mu^\gamma
  k^{\bar\cA\alpha}_{\gamma}I^{n-2}_{\bar\cA}F^{I}\d_I\Theta_{\alpha}
  +P^\Sigma(F)F^I\d_I\Theta_\Sigma+
  dI^{n-1\alpha}\Theta_\alpha\approx 0. 
\end{equation}

\subsubsection{Pure abelian Yang-Mills theory}
\label{sec:pure-abelian-yang}

Abelian Yang-Mills theory is treated in
\cite{Barnich:1993pa,Barnich:1994mt} and in section 13 of
\cite{Barnich:2000zw}. From the current perspective, There are no
$\bar\cA$ indices and all indices $I$ are $a$ indices, because all
characteristic cohomology in degree $n-2$ is covariantizable,
$s\star A^{*I}+d \star F^I=0$. The obstruction equation becomes
$(-)^n\star F^J F^I\d_I
\Theta_J+P^\Sigma(F)^{n-2}F^I\d_I\Theta_\Sigma+dI^{n-1\alpha}\Theta_\alpha\approx
0$. In the $x^\mu$-independent case, this implies
$(-)^n\star F^J F^I\d_I
\Theta_J+P^\Sigma(F)^{n-2}F^I\d_I\Theta_\Sigma=0$ by putting to zero
derivatives of the $F$'s. Then both terms have to vanish separately.
This can be seen by taking an Euler-Lagrange derivative with respect
to $A^I_\mu$ and using the fact that $P^\Sigma(F)^{n-2}F^I$ is a total
derivative.  This implies $\d_{(I}\Theta_{J)}=0$ which gives in turn
$\Theta_I=\d_I\Theta$ (fermionic Poincar\'e lemma). In this case,
there are no more obstructions and the cubic vertex comes from
$ [d^nx C^{*I}\d_I+\star A^{*I}A^J\d_J\d_I+\half \star F^I A^J
A^K\d_K\d_J\d_I]\Theta$ with $\Theta$ in ghost number $3$. There can
be no $P_A$ classes since $P^n\approx dI^{n-1}$ implies $P^n=0$ when
putting to zero all derivatives of the $F^I_{\mu\nu}$.

There is thus only $U$,$B$ or $I$-type cohomology, and the most
general infinitesimal deformation is described by $n$ forms
$U^{0,n}+B^{0,n}+I^n$. With only abelian factors, the $B^{0,n}$ exists
only in odd spacetime dimensions $n$ and are characterized by
completely symmetric tensors of rank $m$ with $n=2m-1$.

\subsubsection{Higher dimensional Yang-Mills-Chern-Simons theory}
\label{subsec:AFFFF}

Abelian Yang-Mills-Chern-Simons theory in dimension $n=2m-1\,$ is
determined by the action
$\int_{M_{2m-1}} \cL^{2m-1}\,$, where
$\cL^{2m-1} = \tfrac{1}{2}\,F^I\star F^J \delta_{IJ}
+\tfrac{1}{m}\,d_{IJ_1\ldots J_{m-1}} A^I F^{J_1}\ldots F^{J_{m-1}}\,$
and $d_{J_1\ldots J_m}=d_{(J_1\ldots J_m)}\,$ is a totally symmetric,
constant symbol.  We assume linear independence of the $n_v$ symmetric
tensors of rank $m-1$ obtained by letting the index $I$ in
$d_{IJ_1\ldots J_{m-1}}\,$ run over its $n_v$ values.  This is
equivalent to requiring that the $d$ symbol does not possess any nul
vector, in the sense that $V^I\,d_{IJ_1\ldots J_{m-1}}=0$
$\Leftrightarrow$ $V^I=0$ for all $I\,$. For $n=3$, $m=2\,$, this
means that $d_{IJ}$ is a nondegenerate bilinear form on
$\mathbb{R}^{n_v}\,$ that we take to be $\delta_{IJ}\,$. Below,
abelian indices are lowered and raised with $\delta_{IJ}$ and its
inverse.

The field equations read
$0\approx \frac{\delta \cL^{2m-1}}{\delta A^I} = d_{IJ(m-1)}F^{J(m-1)}
+d \star F_I\,$, where we use the notation
$d_{IJ(m-1)} = d_{IJ_1\ldots J_{m-1}}$ and
$F^{J(m-1)} = F^{J_1}\ldots F^{J_{m-1}}\,$.  This is equivalent to
$s\star A^*_I + d\star F_I + d_{IJ(m-1)}F^{J(m-1)}=0$. There thus
exist characteristic classes that are trivial in
$H^{n-1}_{\rm char}(d,{\cal I})$:
$P^{n-1}_{\bar {\cal A}} = d_{IJ(m-1)}F^{J(m-1)}\,$.  Indeed, our
assumption on the symbol $d_{IJ_1\ldots J_{m-1}}\,$ implies that there
are exactly $n_v$ independent characteristic classes that are trivial
in $H^{n-1}_{\rm char}(d,{\cal I})\,$, so that the index
$\bar{\cal A}$ can be identified with the index $I$ running over the
$n_v$ abelian factors.  In other words, there is no covariantizable
characteristic cohomology in degree $n-2$, there are no $k^{n-2}_a$,
and all indices $I$ are again $\bar\cA$ indices. Note also that the
invariant $I^{n-2}_{\bar {\cal A}}$ from the definition
$P^{n-1}_{\bar {\cal A}}\approx dI^{n-2}_{\bar {\cal A}}$ is equal to
$-\star F_I\,$. There is thus no $U$-type cohomology. There is no
$W$-type cohomology either since the spacetime dimension is odd.

In order to construct cohomology classes of $R$-type as in
\eqref{C.25}, we need to determine (a basis of) the intersection
$k^{\bar {\cal A}\alpha} P^{n-1}_{\bar {\cal
    A}}\Theta_\alpha=N^{n-1}\,$, where
$N^{n-1}= sb^{n-1} + dB^{n-2}\,$.  From the characterization of
$H(s|d,\cal B)$ in \eqref{B.38}, this intersection is determined by
constants $k^{I\,\alpha}\equiv \frac{1}{k!}f^{I}{}_{J_1\ldots J_k}$ such that
\begin{align}
    \frac{1}{k!}f^{K}{}_{I_1\ldots I_k}\,d_{KJ(m-1)}F^{J(m-1)}C^{I_1}\ldots C^{I_k} = 
  \tfrac{m+k-1}{m-1}\,\lambda_{J(m-1),I_1\ldots I_k}\,
  F^{J(m-1)}C^{I_1}\ldots C^{I_k}\;.\label{relation}
\end{align}
In other words, the constants $f^{K}{}_{I_1\ldots I_k}$
should be such that
$\frac{1}{k!}\,f^{K}{}_{I_1\ldots I_k}\,d_{KJ(m-1)}=
\frac{m+k-1}{m-1}\,\lambda_{J(m-1),I_1\ldots I_k}\,$. The symmetry
\eqref{eq:lamsym} then implies the constraint
\begin{align}
    d_{K (J_{1}\ldots J_{m-1}} f^K{}_{J_{m}) I_{2}\ldots I_{k}} = 0\;.
    \label{constraintonk}
\end{align}
In particular, for $n=3$ with $d_{IJ}=\delta_{IJ}\,$, this requires
$f_{I J_1\ldots J_k}:=\delta_{IL}\,f^{L}{}_{J_1\ldots J_k}$ to be
completely antisymmetric, so that
$k^{\bar{\cal A}\alpha}\Theta_\alpha \equiv \frac{1}{k!}f^{I}{}_{J_1\ldots
  J_k}C^{J_1}\ldots C^{J_k}$ is equal to $\partial^I\Theta$ with
\begin{align}
\Theta=\tfrac{1}{(k+1)!}\,f_{J_1\ldots J_{k+1}}C^{J_1}\ldots
C^{J_{k+1}}\,.\label{4.6}
\end{align}
More generally, this also implies that all $N$'s of form degree
$n-1=2m$ are $N_\gamma$'s for particular constants 
$f^{I}{}_{J_{1}\ldots J_{k}}$ and $d_{IJ_{1}\ldots J_{m-1}}$ such 
that \eqref{constraintonk} is true. Indeed, because a tensor 
$\lambda_{J(m-1),I[k]}$ is an irreducible $GL(n_{v})$ tensor with 
symmetry given by the following Young tableau
\begin{align} 
    \YT{_{1} & _{2} & \ldots  & _{m-1} \\ _{1}\\ \vdots \\ _{k} }\quad 
    \cong 
  \Big(~   \YT{ _{1} & _{2}& \ldots  & _{m-1}}
  ~ \otimes~ 
    \YT{ _{1}\\ \vdots \\ _{k} } ~\Big)
    \qquad \ominus \qquad
  \Big(   ~\YT{ _{1}& _{2} & \ldots  & _{m}}
   ~\otimes~ 
    \YT{ _{2}\\ \vdots \\ _{k} } ~\Big)   \;,
    \label{decomposition}
\end{align}
it can be parametrised as 
\begin{align}
\lambda_{J_{1}J_{2}\ldots J_{m-1},I_{1}\ldots I_{k}} 
&= d_{J_{1}J_{2}\ldots J_{m-1}}\,f_{I_{1}\ldots I_{k}}\;,\qquad{\rm{with}}\qquad 
d_{(J_{1}J_{2}\ldots J_{m-1}}\,f_{J_{m})I_{2}\ldots I_{k}} = 0\;,
\label{4.7}
\end{align}
where the latter constraint expresses the subtraction term on the 
right-hand side of \eqref{decomposition}. 
Therefore, a fortiori an $GL(n_{v})$-irreducible tensor $\lambda_{J(m-1),I[k]}$
(with $k\leqslant n_{v}$) can be written as a sum of terms of the type given 
in \eqref{4.7}:
\begin{align}
\lambda_{J_{1}\ldots J_{m-1},I_{1}\ldots I_{k}} 
&= d_{KJ_{1}\ldots J_{m-1}}\,f^{K}{}_{I_{1}\ldots I_{k}}\;,\qquad{\rm{with}}\qquad 
d_{K(J_{1}\ldots J_{m-1}}\,f^{K}{}_{J_{m})I_{2}\ldots I_{k}} = 0\;,
\end{align}
which is what we wanted to show. 
According to \eqref{C.24} and \eqref{abeliandescent}, this
gives
\begin{align}
 W^{n-1} =& \tfrac{1}{2}\,A^K A^L\partial_{L}\partial_{K}\Big[
         \tfrac{m-1}{k!(m+k-1)}\,d_{M J_{1}\ldots J_{m-2}I_{1}}\,
         f^{M}{}_{I_{2}\ldots I_{k+1}}\,
         F^{J_{1}}\ldots F^{J_{m-2}}\,C^{I_{1}}\ldots C^{I_{k+1}}
         \Big]
         \nonumber \\
         &+ \tfrac{1}{k!}\,f^{K}{}_{I_{1}\ldots I_{k}}\,
         \Big(\star \!F_{K} A^{L}\partial_{L}\, 
         + \star A^{*}_{K}\,\Big) C^{I_{1}}\ldots C^{I_{k}}\,.
         \label{eq:12}
\end{align}
As we explained above, in the 3-dimensional case $n=3$ (i.e., $m=2$), 
the constraint \eqref{constraintonk} can be solved explicitly, which allows 
to rewrite  \eqref{eq:12} in a more compact way:
\begin{align}
  \label{eq:123D}
 W^{2} =& \Big[ ( \tfrac{1}{2}\,A^K A^L + \star F^{K} A^{L})\partial_{L}\partial_{K}
            + \star A^{*K}\partial_{K}\,\Big] \,\Theta\;,\qquad 
            \Theta=\tfrac{1}{(k+1)!}\,f_{J_1\ldots J_{k+1}}C^{J_1}\ldots
C^{J_{k+1}}\;.
\end{align}
In dimension $n=2m-1\,$, we know that $dW^{n-1}+sR^{n} = k^{{\cal A}\alpha} I^{n-2}_{\cal A} F^{I}\partial_{I}\Theta_{\alpha}\,$, with 
\begin{align}
  \label{eq:RCS}
 R^{n} =& 
\tfrac{1}{3!}\, A^{K}A^{L}A^{M}\partial_{M}\partial_{L}\partial_{K}\,\Big[
 \tfrac{m-1}{k! (m+k-1)}\,d_{NJ(m-2)I}\,f^{N}{}_{I[k]}\,
 F^{J_{1}}\ldots F^{J_{m-2}}\,C^{I_{1}}\ldots C^{I_{k+1}} \Big]
  \nonumber\\
 & + \tfrac{1}{k!}\,f^{M}{}_{I_{1}\ldots I_{k}}\,( \star F_{M}\,
 \tfrac{1}{2}\,A^K A^L \partial_{L}\partial_{K}
            + \star A^{*}_{M}\,A^{K}\partial_{K} + \star C^{*}_{M}) 
            C^{I_{1}}\ldots C^{I_{k}} \;.
\end{align}
In dimension $n>3\,$, the obstruction in lifting $W^{n-1}$ is due to the fact that 
$k^{{\cal A}\alpha} I^{n-2}_{\cal A}N^{2}_{\alpha} =  
k^{{\cal A}\alpha} I^{n-2}_{\cal A} F^{I}\partial_{I}\Theta_{\alpha}=
\tfrac{1}{k!}\,f^{K}{}_{I[k]}\,\star F_{K}F^{L}\partial_{L}C^{I[k]}\,$ 
does not vanish because the constant symbol 
$f_{K | J[k]}=\delta_{KM}\,f^{M}{}_{I[k]}$ is not necessarily antisymmetric. 
In dimension $n=3\,$, the constraint \eqref{constraintonk} entails 
antisymmetry of the $f_{K | J[k]}$ and therefore no obstruction arises in 
the lift of the 2-form $W^{2}\,$.

Furthermore, because $n$ is odd, the obstruction equation
\eqref{eq:8}  reduces to 
\begin{equation}
k^{\bar{\cal A}\alpha}I^{n-2}_{\bar{\cal A}} F^{I}\partial_{I}\Theta_{\alpha} 
+ N^{n}_{(\mu)} + dI_{(\mu)}\,\Theta_{\alpha} + s(K_{(\mu)}\Theta_{\alpha}) = 0\;,
\qquad \Theta_{\alpha} = C^{I_{1}}\ldots C^{I_{k}}\;,
%\star F^K F^L\d_L\d_K\lambda_{J(m-2)I,I[k-1]}F^{J(m-2)}C^{I[k]} +d
%I^{n-1\alpha}\Theta_{\alpha}\approx 
%0\label{eq:13}\,.
\end{equation}
or, more explicitly, 
\begin{equation}
-\tfrac{1}{k!}\,f^{K}{}_{I_{1}\ldots I_{k}}\star F_{K} F^{I_{1}} C^{I_{2}}\ldots C^{I_{k}}
+ N^{n}_{(\mu)} + dI_{(\mu)}\,\Theta_{\alpha} + s(K_{(\mu)}\Theta_{\alpha}) = 0\;.
\end{equation}
In the case where the $\mathfrak g\,$-invariant local 
functions in ${\cal I}$ do not explicitly depend on  $x^\mu\,$, 
there is no solution to the obstruction equation if the first term 
does not vanish identically. 
It does vanish if and only if the symbol 
$f_{K | J[k]}=\delta_{KM}\,f^{M}{}_{I[k]}$ is completely antisymmetric, 
on account of $\star F^K F^L=\star F^L F^K\,$. 
In this case, there is no obstruction and no need for
$I^{n-1\alpha}_\mu$ nor $K^{n\alpha}_\mu\,$.
This allows us to write $R^{n}$ and $W^{n-1}$ as follows:
\begin{align}
%  \label{eq:RCSbis}
 R^{n} =& 
\tfrac{1}{3!}\, A^{K}A^{L}A^{M}\partial_{M}\partial_{L}\partial_{K}\,\Big[
 \tfrac{m-1}{k! (m+k-1)}\,d_{NJ(m-2)I}\,f^{N}{}_{I[k]}\,
 F^{J_{1}}\ldots F^{J_{m-2}}\,C^{I_{1}}\ldots C^{I_{k+1}} \Big]
  \nonumber\\
 & + ( \tfrac{1}{2}\star \! F^{K}A^L A^M \partial_{M}\partial_{L}\partial_{K}
            + \star A^{*K}\,A^{L}\,\partial_{L}\,\partial_{K} + \star C^{*K}\partial_{K}) \Theta\;,
        \nonumber \\
 W^{n-1} =& \tfrac{1}{2}\,A^K A^L\partial_{L}\partial_{K}\Big[
         \tfrac{m-1}{k!(m+k-1)}\,d_{M J_{1}\ldots J_{m-2}I_{1}}\,
         f^{M}{}_{I_{2}\ldots I_{k+1}}\,
         F^{J_{1}}\ldots F^{J_{m-2}}\,C^{I_{1}}\ldots C^{I_{k+1}}
         \Big]
         \nonumber \\
         &+ (\star F^{K} A^{L}\,\partial_{L}\partial_{K}\, 
         + \star A^{*K}\,\partial_{K})\Theta\;,
         \qquad \Theta = \tfrac{1}{(k+1)!}\,
            f_{I_{1}\ldots I_{k+1}}\,C^{I_{1}}\ldots C^{I_{k+1}} \;,
\end{align}
with the descent equations $sR^{n}+dW^{n-1}=0\,$,  
$sW^{n-1}+dQ^{n-2}= 0\,$,  $sQ^{n-2}=0\,$, where
\begin{align}
Q^{n-2} = A^{K}\partial_{K}\left( \tfrac{m-1}{k!(m-k+1)}\,d_{MJ(m-2)I_{1}}\,f^{M}{}_{I_{2}\ldots I_{k+1}}
\, F^{J(m-2)}\,C^{I_{1}}\ldots C^{I_{k+1}} \right)+ \star F^{K}\partial_{K}\,\Theta\;.
\end{align}
In particular, in $n=3$ dimensions (i.e. $m=2$), it yields 
\begin{align}
%  \label{eq:RCSbis}
 R^{3} =& \Big( \tfrac{1}{2} [ \star  F^{K}A^L A^M + \tfrac{1}{3}\, A^{K}A^{L}A^{M}]
 \partial_{M}\partial_{L}\partial_{K}
            + \star A^{*K}\,A^{L}\,\partial_{L}\,\partial_{K} + \star C^{*K}\partial_{K}\Big) \Theta\;,
\end{align}
with $W^{2}$ already given in \eqref{eq:123D} and $Q^{1}= (A^{I}+\star F^{I})\,\partial_{I}\Theta\,$.
Note that $R^{3}$ contains the expected cubic vertices for the 3-dimensional non-abelian 
Yang-Mills-Chern-Simons theory.

An R-type infinitesimal deformation requires ghost number $0$ and thus
$k=2$, so that \eqref{relation} reduces to 
\begin{align}
\tfrac{1}{2}\,
f^{K}{}_{I_1I_2}\,d_{KJ(m-1)}=\tfrac{m+1}{m-1}\,\lambda_{J(m-1),I_1I_2}\,,    
\end{align}
the zero ghost number component of $R^{n}$ being 
\begin{align}
    R^{0,2m-1} = &\;\tfrac{1}{2}\,f^J{}_{I_1I_2}\,\star \!C^*_J\, C^{I_1}C^{I_2} + 
                 f^{J}{}_{I_1I_2}\,
                 (\star A^*_J \,A^{I_1} C^{I_2}+\tfrac{1}{2}\star\! F_JA^{I_1}A^{I_2}) 
    \nonumber \\
    & + \;\tfrac{m-1}{m+1}\,\tfrac{1}{2}\,f^{K}{}_{I_1I_2}\,d_{I_3 J(m-2)K}\,
    F^{J(m-2)}\,A^{I_1}A^{I_2}A^{I_3} \;.
    \label{defoCSodd}
\end{align}
As before, when asking for the infinitesimal deformation $S^{(1)}$ to
be consistent at second order,
$\half (S^{(1)},S^{(1)})+(S^{(0)},S^{(2)})=0\,$, one finds the Jacobi
identity for $f^K{}_{IJ}$ by asking that the term in $C^*_{I}$, which
is the tail of a non trivial cohomology class, be absent. There are no
further conditions. The constraint \eqref{constraintonk} can then be
interpreted as an invariance condition for the totally symmetric $d$
symbol under the Lie algebra $\mathfrak g$ with structure constants
$f^K{}_{IJ}\,$. The complete deformation does not require further
antifield dependent terms. The additional antifield independent terms
add up to give the non-abelian Yang-Mills action in dimension $n=2m-1$
plus the standard non-abelian Chern-Simons Lagrangian given by the
homotopy integral
\begin{align}
    \cL^{2m-1}_{na} = \int_0^1 dt\,d_{I(m)}\left[ A^I(dA+t^2A^2)^{I(m-1)}
    \right]\;.
\end{align}

\subsubsection{Axion models in even dimensions}
\label{subsec:Peccei-Quinn}

We now consider even spacetime dimension $n=2m$ with Lagrangian
\begin{align}
    {\cal L} =   - \frac{1}{2}\, \partial_\mu \phi^I \partial^\mu \phi^J\, \delta_{IJ} 
    +\frac{1}{2}\,{\delta}_{IJ} F^I \star\! F^J + \phi^I\,
    t_{I|J_1\ldots J_m}\,F^{J_1}\ldots F^{J_m}\;,
\end{align}
for some constants $t_{I,J_1\ldots J_m}\,$, and concentrate on $W$-type
cohomology classes.

The field equations for the $n_s=n_v$ scalars are responsible for the
existence of characteristic classes
$P^n_{I} = t_{I|J_1\ldots J_m}\,F^{J_1}\ldots F^{J_m}$ such that
$P_I = dI^{n-1}_I - sK_I\,$ where $K_I=\star \phi^*_I\,$.  In order to
find $W$-type cohomology classes as in \eqref{C.29},
$W = b^n - k^{I \alpha} (K_I\Theta_\alpha +
I^{n-1}_I[\Theta_\alpha]^1)\,$, we need to determine the intersection
$sb^n + d b^{n-1} = N_{m,k} =k^{A\alpha} P_A\Theta_\alpha$ where
$N = \tfrac{m+k}{m}\,\lambda_{J(m),I[k]}$
$F^{J_1}\ldots F^{J_m}\,C^{I_1}\ldots C^{I_k}\,$,
$b^n=\tfrac{1}{2}\,A^{K}A^{L}\,\partial_{L}\partial_{K}\,
(\lambda_{J(m-1)I,I[k]}\,F^{J(m-1)}C^{I[k+1]})\,$
and $I^{n-1}_I = \star d\phi_I\,$.
The intersection condition is satisfied provided
\begin{align}
    \tfrac{m+k}{m}\,\lambda_{J_{1}\ldots J_{m},I_{1}\ldots I_{k}} = 
    t_{K|J_{1}\ldots J_{m}} \, k^{K}{}_{I_{1}\ldots I_{k}} \;.
    \label{4.20}
\end{align}
The symmetry properties of $\lambda_{J(m),I[k]}$ imply the following
constraint: $t_{K|(J_{1}\ldots J_{m}} \, k^{K}{}_{J_{m+1})I_{2}\ldots I_{k}} =0\,$. 
Similarly to the analysis leading to \eqref{4.7}, any $GL(n_{v})$-irreducible tensor 
$\lambda_{J_{1}\ldots J_{m}, I_{1}\ldots I_{k}}$ with $k\leqslant n_{v}$ 
can be written as in \eqref{4.20} provided  the latter constraint is fulfilled. 

This condition is fulfilled by taking
$k_{L|I[k]}=\delta_{KL}\,k^K{}_{I[k]} = f_{LI_{1}\ldots I_{k}}\,$, for 
$f_{LI_{1}\ldots I_{k}}$ an antisymmetric tensor of rank $k+1$ and
$t_{I|J(m)} = \delta_{IJ}\check{t}_{J(m-1)}\,$ for 
$\check{t}_{J(m-1)}$ a symmetric rank-$(m-1)$ tensor $\check{t}\,$. 
Introducing
$\Theta = \tfrac{1}{k+1}\,f_{I_{1}\ldots I_{k+1}}C^{I_{1}}\ldots C^{I_{k+1}}\,$,
the W-type cohomology classes are then given by
\begin{align}
    W^{k-1,2m} =& \tfrac{1}{2(m+k)}A^{K}A^{L}\,\partial_{L}\partial_{K} \left[
    \check{t}_{J(m-1)}\,f_{I[k+1]} + (-)^k \,(m-1) \,\check{t}_{J(m-2)I}\,f_{I[k]J}
    \right] F^{J(m-2)}C^{I_{1}}\ldots C^{I_{k+1}}
    \nonumber \\
    & - (\star \phi^{*J}\,\partial_{J} + \star d\phi^J  A^{K}\,\partial_{K}\partial_{J} )\Theta\;.
\end{align}
Infinitesimal deformations require ghost
number $0$ and thus $k=1$. The first term on the right-hand side then
exactly reproduces the first order deformation in the Lagrangian (35) of
\cite{deWit:1987ph}, provided one identifies the constants
$C_{I,J_1\ldots J_m}$ therein with $\lambda_{J_1\ldots J_m,I}\,$.

The examples of this section generalise to arbitrary even dimensions
the $W$-type cohomology classes in 4 spacetime dimensions
discussed in section 6.4 of \cite{Barnich:2017nty}.

\subsection{Coupled abelian vector-scalar models in four
  dimensions}

The results for this case are discussed in detail in section 3.2 of
\cite{Barnich:2017nty}.

\subsection*{Acknowledgments}

\addcontentsline{toc}{section}{Acknowledgments}

The authors thank M.~Henneaux, B.~Julia, V.~Lekeu and A.~Ranjbar for
useful discussions. GB is grateful to F.~Brandt for earlier
collaborations on this subject. This work was partially supported 
by FNRS-Belgium (convention FRFC PDR T.1025.14 and convention IISN
4.4503.15).  NB is Senior Research Associate of the F.R.S.-FNRS.

\appendix

\section{Conventions}
\label{sec:notations}

The components of the Minkowski metric are given, in inertial 
coordinates in which we work, by the mostly plus expression 
$\eta_{\mu\nu}={\rm diag}(-1,+1,\dots,+1)\,$. 
The symbol $\epsilon_{\mu_1\dots\mu_n}$ denotes the completely 
antisymmetric Levi-Civita density with the convention that 
$\epsilon^{01\dots n-1}=1\,$ so that $\epsilon_{01\dots n-1}=-1\,$. 
A local basis of anticommuting exterior differential 1-forms 
is given by the family $(dx^\mu)_{\mu = 0, \ldots, n-1\,}$. 
The wedge product symbol $\wedge$ will always be omitted.   

We will sometimes use the notation 
$(d^{ n - p }x)_{\mu_{n-p+1}\dots\mu_n} := -\frac{1}{p!(n -
  p)!}dx^{\mu_{1}}\dots dx^{\mu_{n-p}}\epsilon_{\mu_1\dots\mu_n}$
for $1\leqslant p \leqslant n\,$, and $d^nx:=dx^0\dots dx^{n-1}\,$.
The Hodge dual of a differential $p\,$-form
$\omega^p \equiv \frac{1}{p!}\,dx^{\mu_1}\dots
dx^{\mu_p}\omega_{\mu_1\dots\mu_p}$,
is the $n-p\,$-form given, in our convention, by 
\begin{align}
\star\,\omega^p&=\tfrac{1}{p!(n-p)!}\,dx^{\nu_1}\dots
dx^{\nu_{n-p}}\epsilon_{\nu_1\dots\nu_{n-p}\mu_{n-p+1}\ldots\mu_n}
\omega^{\mu_{n-p+1}\dots\mu_n}=-
(d^{ n - p }x)_{\mu_{n-p+1}\dots\mu_n}
\omega^{\mu_{n-p+1}\dots\mu_n}\;. 
\nonumber
\end{align}
As a consequence, the exterior differential of the dual of a 
$p\,$-form reads 
\begin{align}
d \star \omega^p =-(-)^{n-p} (d^{n-p+1}x)_{\nu_1\dots\nu_{p-1}}
\,\partial_{\mu}\omega^{\mu\nu_1\dots\nu_{p-1}}\;.
\end{align}

\section{Proof of the main theorem}
\label{sec:proof}

The proof proceeds by induction on the form degree.  At $p=0\,$ and
due to our assumption $n>2\,$, items (i) and (ii) reduce respectively
to
$s\omega^0 = 0 \Leftrightarrow \omega^0 = I^{0\alpha}\Theta_\alpha + s
\eta^0$ and $I^{0\alpha}\Theta_\alpha = s\eta^0$ $\Leftrightarrow$
$I^{0\alpha}\approx 0\,$. Both hold on account of corollary 11.2 
of \cite{Barnich:2000zw}.  Item (iii) has no content
for $p=0$. Item (iv) holds: if a constant vanishes weakly it vanishes
by assumption on the equations of motion.

We will further decompose the induction into three main steps.  In
step [A], we consider $p=m-1<n-2\,$ and go to $p=m<n-1\,$.  For step
[B] ,we consider the induction going from $p=n-2$ to $p=n-1\,$ and in
step [C] we will finally reach $p=n\,$.

\noindent \textbf{Proof}

\noindent A. Assume now that (i), (ii), (iii), (iv) hold at form
degrees $p=m-1<n-2\,$ and let us show that they hold at form degree
$m<n-1\,$. In this case, the proof proceeds exactly like in theorem
11.1 of \cite{Barnich:2000zw}:

\begin{enumerate}

\item[(iv)] $P^m-dI^m\approx 0$ for $p=m<n-1$ 
  implies that $sK^m + dI^{m-1}=P^m\,$. Combining this with 
  $P^m=dB^{m-1}$ ($B^{m-1}\equiv q^{m-1}$ the Chern-Simons form that 
  we associate with $P^m$) yields $sK^m + d(I^{m-1}-B^{m-1})=0$ 
  hence $[K^m]\in H^{-1,m}(s|d)\,$. The latter group is isomorphic 
  to $H^{m-1}_{\rm char}/\delta^{m-1}_0\mathbb R$ which vanishes for
  $m-1<n-2$, and thus for $m<n-1\,$. 
  Hence, $K^m \sim 0 $ which implies that $I^{m-1}-B^{m-1}\sim 0$
  by the usual properties of descent equations. 
  In turn, this implies $B^{m-1}=0$ on account of (ii) 
  that is supposed to 
  hold for $p=m-1\,$, and thus $P^m(F)=0\,$. 
  This proves (iv) for $p=m<n-1\,$.

\item[(i)] The cocycle condition $s\omega^m+d\omega^{m-1}=0$
  implies that $\omega^{m-1}$ satisfies the similar equation
  $s\omega^{m-1}+d\omega^{m-2}=0$ and thus that
  \begin{align}
    \omega^{m-1}=I^{m-1\alpha}\Theta_\alpha+{B}^{m-1}  
  \end{align}
  since we assume
  that (i) holds at degree $m-1<n-2$ and since trivial contributions
  can be neglected. This implies
  $d\omega^{m-1}=dI^{m-1\alpha}\Theta_\alpha
  -s(I^{m-1\alpha}[\Theta_\alpha]^1+ \hat B^m + b^m)+N^m\,$, 
  by virtue of the relation $s[\Theta_\alpha]{}^1 + d \Theta_\alpha =
  0\,$, and the relation \eqref{11.37},
  $d{B}^{m-1} = -s(\hat B^m + b^m)+N^m$, for
  some $\hat B^m,b^m,N^m$ . 

  When inserted in the equation for $\omega^m$ and using 
  corollary 11.2 of \cite{Barnich:2000zw}, we find
  that $d I^{m-1\alpha}+P^{m\alpha}(F)\approx 0\,$, where
  $N^m=P^{m\alpha}\Theta_\alpha\,$. Since we have already shown that
  (iv) holds for $m < n-1$, we get that $P^{m\alpha}=0=N^m=b^m\,$.  
  This implies $d{B}^{m-1}+s\hat B^{m}=0$ and 
  $d I^{m-1\alpha}\approx 0$
  which in turn implies $I^{m-1\alpha}\approx d\omega^{m-2\alpha}$ 
  (or constant for $m-1=0$) since there is no non-trivial characteristic
  cohomology in degree $m-1<n-2\,$. 
  Since (iii) is assumed to hold for $p=m-1\,$, we find
  $I^{m-1\,\alpha}\approx dI^{m-2\,\alpha}+P^{m-1\,\alpha}\,$. 
  By using corollary 11.1 of \cite{Barnich:2000zw}, this yields 
  \begin{multline}
    I^{m-1\,\alpha}=sK^{m-1\,\alpha} + 
    dI^{m-2\,\alpha}+P^{m-1\,\alpha} \\ \Longrightarrow
    \omega^{m-1}=[P^{m-1\,\alpha}+dI^{m-2\,\alpha}+sK^{m-1\,\alpha}]
    \Theta_\alpha+{B}^{m-1}. 
    \label{Im-1alpha}
  \end{multline}
  The part $P^{m-1\alpha}\Theta_\alpha$ can be decomposed in terms 
  of the basis of $H(s,{\cal B})$ consisting of the $M$'s and $N$'s:
  $P^{m-1\alpha}\Theta_\alpha=N^{m-1}+M^{m-1}+\delta^{m-1}_0
  \hat\lambda\,$. Neither the constant nor $N^{m-1}\sim 0$ will
  contribute when used to calculate $\omega^m\,$.  Since
  $[dI^{m-2\alpha}+sK^{m-1\alpha}]\Theta_\alpha
  =d(I^{m-2\alpha}\Theta_\alpha)+s(
  I^{m-2\alpha}[\Theta_\alpha]^1+K^{m-1\alpha}\Theta_\alpha)$, these
  terms will not contribute either. We can thus assume that
  $\omega^{m-1}=M^{m-1}+{B}^{m-1}\,$.  Since $dM^{m-1}=-s\bar B^{m}\,$
  for some $\bar B^{m}\in\cal B\,$, if we denote
  $B^m := \bar B^m+\hat B^m$, we have shown that
  $d\omega^{m-1}+sB^m=0\,$. Finally, the initial cocycle condition
  $s\omega^m + d\omega^{m-1}=0$ now reduces to $s(\omega^m-B^m)=0$ and
  thus $\omega^m=I^{m\alpha}\Theta_\alpha+B^m+s\eta^m\,$ by 
  corollary 11.2 of \cite{Barnich:2000zw}.  Taking into account the
  trivial terms that we have neglected (in particular $N^{m-1}$), we
  find $\omega^m\sim I^{m\alpha}\Theta_\alpha+B^m\,$, which is (i) at
  form degree $m<n-1\,$.

\item[(ii)] Consider $\omega^m=I^{m\alpha}\Theta_\alpha+B^m\,$
  (we recall that $m<n-1$) where $sB^m+d(B^{m-1}+M^{m-1})=0\,$. The
  cocycle condition $s\omega^m+d\omega^{m-1}=0$ is satisfied with
  $\omega^{m-1}={B}^{m-1}+I^{m-1\alpha}\Theta_\alpha\,$ and where
  $I^{m-1\alpha}\Theta_\alpha = M^{m-1}\,$.  Assume now
  that $\omega^m=I^{m\alpha}\Theta_\alpha+B^m\sim 0\,$.  This implies
  that
  $\omega^{m-1}={B}^{m-1}+I^{m-1\alpha}\Theta_\alpha\sim 0\,$.
  Since we assume that (ii) holds at form degree $m-1\,$, we find
  ${B}^{m-1}=0\,$,
  $M^{m-1}=I^{m-1\alpha}\Theta_\alpha\approx
  N^{m-1}+dI^{m-2\,\alpha}\Theta_\alpha\,$. This implies
  $P^{m-1\alpha}\Theta_\alpha:= M^{m-1}-N^{m-1}\approx
  dI^{m-2\,\alpha}\Theta_\alpha\,$, which gives on account of (iv)
  that $P^{m-1\alpha}=0$ and thus that $M^{m-1}=N^{m-1}$ which means
  that both have to vanish separately, since they are independent
  elements of the basis of $H(s,\cal B)\,$. It then follows that $sB^m=0$, which
  implies that $B^m=0$ by definition of $B^m\,$. We thus find
  $I^{m\alpha}\Theta_\alpha=s\eta^m+d\eta^{m-1}\,$.  Applying $s$ to
  this equation gives $s\eta^{m-1}+d\eta^{m-2}=0$ and then, on account
  of (i), that
  $\eta^{m-1}= \tilde B^{m-1}+\tilde I^{m-1\alpha}\Theta_\alpha +
  s\tilde\eta^{m-1}+d\tilde\eta^{m-2}\,$.  This yields
  $I^{m\alpha}\Theta_\alpha=s(\eta^m-d\tilde\eta^{m-1}-\tilde
  B^m-\tilde b^m-\tilde I^{m-1\alpha}[\Theta_\alpha]^1)+\tilde N^m+d\tilde
  I^{m-1\alpha}\Theta_\alpha$, which implies
  $I^{m\alpha}-\tilde P^{m\alpha}-d\tilde I^{m-1\alpha}\approx 0$
  where $\tilde N^m=\tilde P^{m\alpha}\Theta_\alpha\,$. Together with
  the relation $B^m=0$ obtained above, this shows (ii) at form degree
  $p=m<n-1\,$.

\item[(iii)]  $I^m\approx d\omega^{m-1}$ implies $I^m\sim 0\,$. Using
  (ii) at $m$, which we have proved, gives $I^m\approx
  dI^{m-1}+P^m$. The converse is trivial since $P^m=dq^{m-1}$.
\end{enumerate}

\noindent B. Let us now proceed to form degree $p=n-1\,$.

\begin{enumerate}

\item[(iv)] For $p=n-1\,$, there is nothing to be proved. Indeed, the
  condition $dq^{n-2}=P^{n-1}(F)\approx dI^{n-2}$ means that
  $I^{n-2}-q^{n-2}$ is an element of characteristic cohomology in form
  degree $n-2\,$.  But we have assumed that the set of
  $P^{n-1}_{\bar{\cal A}}$'s form a basis of characteristic classes
  that are weakly $d\,$-exact in $\cal I\,$,
  $P^{n-1}=\lambda^{\bar {\cal A}}\,P^{n-1}_{\bar{\cal A}}$. For later
  use, we note that it follows from equation \eqref{C.20} that
  $P^{n-1}=\lambda^{\bar\cA}(dI^{n-2}_{\bar\cA}-s\star
  A^*_{\bar\cA})\,$.

\item[(i)] Following the same proof as before when $p<n-1\,$, we
  get from $s\omega^{n-1}+d\omega^{n-2}=0$ that
  \begin{align}
    \omega^{n-2} = I^{n-2\,\alpha}\Theta_\alpha + {B}^{n-2}  
    \label{C.38}
  \end{align}
  with $d {B}^{n-2} = -s(\hat{B}^{n-1}+b^{n-1})+N^{n-1}$ for some
  $\hat{B}^{n-1},b^{n-1}$ and $N^{n-1}=P^{n-1\,\alpha} \Theta_\alpha$,
  which is the obstruction to the lift of an element in $\cal B\,$. 
  The polynomial in the curvatures 
  $P^{n-1\,\alpha}=P^{n-1\,\alpha}(F)$ are themselves also
  obstructions to a lift in the small algebra, being a 
  linear combination of the $N$'s only --- the $M$'s  
  explicitly depend on the undifferentiated ghosts. 
  Computing $d\omega^{n-2}$ and plugging back into the 
  initial cocycle relation, we find 
  \begin{align}
   dI^{n-2\,\alpha}+P^{n-1\alpha}(F)\approx 0\;,  
   \label{C.39}
  \end{align}
  where Lemma 1 was used, as before.  It follows from (iv) at $p=n-1$
  that
  $P^{n-1\alpha} = \lambda^{\bar\cA\alpha} P^{n-1}_{\bar\cA}\approx
  \lambda^{\bar\cA\alpha} dI^{n-2}_{\bar\cA}\,$.  Combining this with
  relation \eqref{C.39} leads to
  $d(I^{n-2\,\alpha}+\lambda^{\bar {\cal A}\alpha} I^{n-2}_{\bar {\cal
      A}})\approx 0\,$. This means that
  $I^{n-2\,\alpha}+\lambda^{\bar {\cal A}\alpha} I^{n-2}_{\bar {\cal
      A}}$ defines a covariant element of characteristic cohomology in
  form degree $n-2\,$.  Therefore
  $I^{n-2\,\alpha}+\lambda^{{\bar\cA}\alpha}I^{n-2}_{\bar\cA}\approx
  \lambda^{a\alpha}k^{n-2}_a+d\omega^{n-3\alpha}\,$, when taking into
  account the definitions before the descent equations \eqref{C.11}.
  Using (iii) for $p=n-2\,$ yields
  \begin{align}
    I^{n-2\,\alpha}+\lambda^{\bar \cA\alpha}I^{n-2}_{\bar \cA}-
  \lambda^{a\alpha}k^{n-2}_a & =P^{n-2\,\alpha}(F) +dI^{n-3\,\alpha}+s
  K^{n-2\alpha} \;,
  \label{C.40}
  \end{align}
  where corollary 11.1 of \cite{Barnich:2000zw} has been used. 

  From the definition of  
  $N_\gamma =k^{\bar\cA\,\alpha}_{\gamma}P^{n-1}_{\bar\cA}\Theta_\alpha$ in item 4 of 
  the list of Ingredients and the fact that 
  $N^{n-1} = P^{n-1\,\alpha}\Theta_\alpha = \lambda^{\bar{\cal
      A}\alpha}\,P^{n-1}_{\bar{\cal A}}\Theta_\alpha\,$,  
  we can decompose $N^{n-1}=\lambda^\gamma N_\gamma\,$ and   
  identify 
  \begin{align}
    \lambda^\gamma k_{\gamma}^{\bar\cA\alpha}
  =\lambda^{\bar\cA\alpha}\;.  \label{C.41}
  \end{align}
  Consider then
  \begin{align}
    \hat\omega^{n-1}&:=\omega^{n-1}-\lambda^{a\alpha}T_{a\alpha}
                          -\lambda^\gamma W_\gamma  \;,\nonumber\\
    \hat\omega^{n-2}&:=\omega^{n-2}
    -\lambda^{a\alpha}k^{n-2}_a\Theta_\alpha 
    - \lambda^\gamma 
    (B^{n-2}_\gamma-k_\gamma^{\bar\cA\alpha}I_{\bar\cA}^{n-2}\Theta_\alpha)\;.
    \label{C.42}
  \end{align}
  They are related by the cocycle relation 
  $s\hat\omega^{n-1}+d\hat\omega^{n-2}=0\,$, as can be seen from 
  \eqref{C.16} and \eqref{C.27}. 
  By using \eqref{C.38} and \eqref{C.40}, we find that 
  \begin{align}
    \hat\omega^{n-2}=[P^{n-2\,\alpha}+dI^{n-3\,\alpha}
     + sK^{n-2\,\alpha}]\Theta_\alpha
     + ({B}^{n-2}-\lambda^\gamma B_\gamma^{n-2})  \;.
  \end{align}
  Following the same reasoning as the one used after the equation 
  \eqref{Im-1alpha}, this then implies 
  $\hat\omega^{n-1}\sim {B}^{n-1}+ I^{n-1\alpha}\Theta_\alpha$ 
  and therefore proves (i) for $p=n-1\,$.

\item[(ii)] Consider
  $\omega^{n-1}=I^{n-1\alpha}\Theta_\alpha+B^{n-1}
  +\lambda^{a\alpha}T_{a\alpha}+\lambda^\gamma W_\gamma$ with
  $s\omega^{n-1}+d\omega^{n-2}=0$ and
  $\omega^{n-2}=\hat B^{n-2}+\hat I^{n-2\alpha}\Theta_\alpha$, with
  $\hat B^{n-2}=B^{n-2}+\lambda^\gamma B_\gamma^{n-2}$, and
  $\hat
  I^{n-2\alpha}\Theta_\alpha=M^{n-2}+\lambda^{a\alpha}k^{n-2}_a\Theta_\alpha-\lambda^\gamma
  k^{\bar\cA\alpha}_\gamma I^{n-2\alpha}_{\bar\cA}\Theta_\alpha$. The
  assumption that $\omega^{n-1}$ is trivial implies that so is
  $\omega^{n-2}$, $\omega^{n-2}\sim 0$. We have shown that (ii) holds
  at form degree $n-2$, so we get
  $B^{n-2}=-\lambda^\gamma B_\gamma^{n-2}$, and
  $\hat I^{n-2\alpha}\Theta_\alpha\approx
  N^{n-2}+dI^{n-3\alpha}\Theta_\alpha$. Since $N_\gamma$ exists only
  if the form degree $n-1$ is even and thus $n$ is odd, there is no
  $N_\gamma$, nor $B_\gamma^{n-2}$ for $n$ even, and thus both
  $B^{n-2}=0$ and $B^{n-2}_\gamma=0$ for $n$ even. For $n$ odd,
  $M^{n-2}=0$. We also have
  $d (\lambda^\gamma B^{n-2}_\gamma)+s(\lambda^\gamma
  b^{n-1}_\gamma)=\lambda^\gamma N_\gamma$. Using this, we find that
  $s (B^{n-1}+\lambda^\gamma b^{n-1}_\gamma)+d(M^{n-2}+
  B^{n-2}+\lambda^\gamma B_\gamma^{n-2})=\lambda^\gamma
  N_\gamma$. Since the $d$ exact term on the left hand side vanishes,  
  we get that $\lambda^\gamma=0$ since $\lambda^\gamma N_\gamma$ is
  $s$ exact only if $\lambda_\gamma$ vanishes. This then also shows
  that $B^{n-2}=0$ if $n$ is odd. This then implies that
  $M^{n-2}+\lambda^{a\alpha}k_a^{n-2}\Theta_\alpha \approx
  N^{n-2}+dI^{n-3\alpha}\Theta_\alpha$. Since
  $M^{n-2}-N^{n-2}=P^{n-2\alpha}\Theta_\alpha=dq^{n-3\alpha}\Theta_\alpha$,
  we find $\lambda^{a\alpha}k_a^{n-2}\approx d\omega^{n-3}$ which
  implies $\lambda^{a\alpha}=0$ since they represent non trivial
  characteristic cohomology classes. We then remain with
  $P^{n-2\alpha}\approx dI^{n-3\alpha}$. Using (iv) at $p=n-2$ then
  gives $P^{n-2\alpha}=0$ and then that $M^{n-2}=0=N^{n-2}$. It
  follows that $sB^{n-1}=0$ which implies $B^{n-1}=0$ since $B^{n-1}$
  needs to have a non trivial descent. We remain with
  $I^{n-1\alpha}\Theta_\alpha=s\eta^{n-1}+d\eta^{n-2}$ and the rest of
  the proof goes through as in the proof of (ii) before since one only
  has to use (i) at form degree $n-2$ which is of the same form than
  at lower form degrees.

\item[(iii)] $I^{n-1}\approx d\omega^{n-1}$ implies $I^{n-1}\sim
  0$. Using (ii) at $p=n-1$, which we have already shown,  gives
  $I^{n-1}\approx dI^{n-2}+P^{n-1}$. The converse obviously holds
  since $P^{n-1}=dq^{n-2}$. 

\end{enumerate}

\noindent C. Finally, let us conclude by dealing with form degree $p=n$.

\begin{enumerate}

\item[(iv)] For $p=n$, there is again nothing to be proved. Indeed, we
  now find that $I^{n-1}-q^{n-1}$ is equivalent to an element of
  non-covariantizable characteristic cohomology in degree $n-1$. But
  we have assumed that the set of $P_{A}$'s form a basis of
  characteristic classes that are weakly $d\,$-exact in $\cal I\,$, so
  that $P^n=\lambda^A P_A$. For later use, we now note that
  $P^{n}=\lambda^Adq^{n-1}_A=\lambda^{A}(dI^{n-1}_{A}+sK_{A})$.

\item[(i)] We can assume in $s\omega^n+d\omega^{n-1}=0$ 
  that $\omega^{n-1}$ is of the form 
  \begin{equation*}
    \omega^{n-1}=I^{n-1\alpha}\Theta_\alpha + 
    B^{n-1}+\lambda^{a\beta}T_{a\beta}+\lambda^\gamma
    W_\gamma\;,
  \end{equation*}
  with $d B^{n-1} = - s(\hat B^{n} + b^{n}) + N^n$, for some $\hat
  B^n, b^n$ and
  $N^n = P^{n\alpha}\Theta_\alpha$.  We then get
  $d\omega^{n-1}=(dI^{n-1\alpha})\Theta_\alpha
  -s(I^{n-1\alpha}[\Theta_\alpha]^1+\hat B^{n}+\hat b^n
  +\lambda^{a\beta}U_{a\beta} +\lambda^\gamma
  R_\gamma)+N^n+[\lambda^{a\beta}(-)^{n}k^{n-2}_a
  +\lambda^{\gamma}k_\gamma^{\bar\cA\beta}
  I^{n-2}_{\bar\cA}]N^2_\beta\,$.  Inserting this into
  $s\omega^n+d\omega^{n-1}=0$ gives
\begin{multline*}
  \left(dI^{n-1\alpha}+P^{n\alpha}+[\lambda^{a\beta}(-)^{n}k^{n-2}_a
  +\lambda^{\gamma}k_\gamma^{\bar\cA\beta}I^{n-2}_{\bar\cA}]P^{2\alpha}_\beta\right)
  \Theta_\alpha \\ +s(\omega^n-I^{n-1\alpha}[\Theta_\alpha]^1
  -\hat B^{n}-b^n-\lambda^{a\alpha}U_{a\alpha}-\lambda^\gamma
  R_\gamma)=0
\end{multline*}
with  
  $N^2_\alpha = P_\alpha^{2\,\beta}\,\Theta_\beta\,$.
  This implies
  \begin{align}
    P^{n\alpha}+dI^{n-1\alpha} + 
  [\lambda^{a\beta}(-)^{n}k^{n-2}_a
  +\lambda^{\gamma}k_\gamma^{\bar\cA\beta}I^{n-2}_{\bar\cA}]
  P^{2\alpha}_\beta \approx 0\;.  \label{eq:ob}
  \end{align}
  This equation is equivalent to the obstruction equation
  \eqref{eq:5}, formulated without antifields and ghosts. Indeed, the
  weak equality can be replaced by a strong equality with zero on the
  right-hand side replaced by $s(-K^{n\alpha})$ with $K^{n\alpha}$
  depending linearly on undifferentiated antifields and such that
  $\gamma K^{n\alpha}=0$, --- the required modification of
  $I^{n-1\alpha}$ corresponds to a trivial term in
  $\omega^{n-1}\,$. Those $\lambda^{a\beta},\lambda^{\gamma}$ for
  which one cannot find a solution to this equation correspond to
  parts of $\omega^{n-1}$ that cannot give a solution to
  $s\omega^n+d\omega^{n-1}=0\,$. Let us denote by
  $\mu^{a\beta},\mu^\gamma$ the general solution with non-vanishing
  first term and particular $P^{n\alpha}_\mu\,$ and
  $I^{n-1\alpha}_\mu\,$ solving this equation. It means that both
  these quantities vanish when $\mu^{a\beta}$ and $\mu^\gamma$ vanish.
  We can thus assume that
  \begin{align}
    \omega^{n-1}&=
                  (I^{n-1\alpha}_R+I^{n-1\alpha}_\mu)\Theta_\alpha 
                  +  B^{n-1}_R +  B^{n-1}_\mu +\mu^{a\beta}T^{n-1}_{a\beta}
                  + \mu^\gamma W^{n-1}_\gamma, \nonumber 
  \end{align}
  and have
  \begin{align}
    P^{n\alpha}_\mu+dI^{n-1\alpha}_\mu + [\mu^{a\beta}(-)^{n}k^{n-2}_a
  +\mu^{\gamma}k_\gamma^{\bar\cA\beta}I^{n-2}_{\bar\cA}]
  P^{2\alpha}_\beta +s K^{n\alpha}_\mu=0\;. \nonumber
  \end{align}
  Upon multiplying by $\Theta_\alpha\,$, this yields
  \begin{align}
    [\mu^{a\beta}(-)^{n}k^{n-2}_a
  +\mu^{\gamma}k_\gamma^{\bar\cA\beta}I^{n-2}_{\bar\cA}]N^{2}_\beta
  + N^n_\mu+d(I^{n-1\alpha}_\mu\Theta_\alpha)
  +s( K^{n\alpha}_\mu\Theta_\alpha+I^{n-1\alpha}_\mu[\Theta_\alpha]^1)
  =0\;.  \nonumber
  \end{align}
  We then define
  \begin{align}
    \bar\omega^{n-1}&:= \omega^{n-1}
     -I^{n-1\alpha}_\mu\Theta_\alpha -  B^{n-1}_\mu
     - \mu^{a\beta}T^{n-1}_{a\beta}-\mu^\gamma W^{n-1}_\gamma
                      \equiv  I^{n-1\alpha}_R\Theta_\alpha +  B^{n-1}_R\;,
    \nonumber\\
    \bar \omega^n &= \omega^n-I^{n-1\alpha}_\mu[\Theta_\alpha]^1
    -\hat B^{n}_\mu -  b^n_\mu -\mu^{a\alpha}U_{a\alpha}
    -\mu^\gamma R_\gamma-K^{n\alpha}_\mu\Theta_\alpha  \;,
    \nonumber
  \end{align}
  where 
  $d B_R^{n-1} = - s( \hat B_R^{n} + b_R^{n}) + N_R^n\,$
  and we denote $N_R^n = P^{n\alpha}_R\Theta_\alpha\,$.
  Similarly, we have that
  $d B_\mu^{n-1} = - s( \hat B_\mu^{n} +  b_\mu^{n}) 
  + N_\mu^n\,$
  with $N_\mu^n = P^{n\alpha}_\mu\Theta_\alpha\,$. 
  The cochains $\bar \omega^n$ and $\bar \omega^{n-1}$ 
  satisfy $s\bar \omega^n+d\bar \omega^{n-1} = 0\,$.
  We then get
  $s\bar\omega^n+d(I^{n-1\alpha}_R\Theta_\alpha+B^{n-1}_R)=0\,$, 
  which gives
  \begin{align}
    s(\bar\omega^n-I^{n-1\alpha}_R[\Theta_\alpha]^1
    -\hat B^n_R-b^n_R)+(dI^{n-1\alpha}_R+P^{n\alpha}_R)\Theta_\alpha  = 0 \; \Longrightarrow
    P^{n\alpha}_R+dI^{n-1\alpha}_R & \approx 0\;. 
  \end{align}
  It follows from (iv) in form degree $p=n$ that $P^{n\alpha}_R=\lambda^{A\alpha}_R P_A$ and then
  $d(I^{n-1\alpha}_R+\lambda^{A\alpha}_R I^{n-1\alpha}_A)\approx 0$ so
  that
  $I^{n-1\alpha}_R+\lambda^{A\alpha}_R I^{n-1\alpha}_A\approx
  \lambda^{\Delta\alpha}j_\Delta+d\omega^{n-2}\,$. 
  The property (iii) at $p=n-1$ then gives
  \begin{align}
    I^{n-1\alpha}_R=-\lambda^{A\alpha}_R\,I^{n-1\alpha}_A+
  \lambda^{\Delta\alpha}j_\Delta+P^{n-1\alpha}_R
  +dI^{n-2\alpha}_R+sK^{n-1\alpha}_R\;. 
  \label{C.60}
  \end{align}
  On the other hand, since
  $N^n_R=P^{n\alpha}_R\Theta_\alpha$ is an obstruction to the 
  lift of an element in $\cal B\,$ and since 
  $P^{n\alpha}_R = \lambda^{A\alpha}_R P_A
  \approx\lambda^{A\alpha}_R dI^{n-1}_A\,$, 
  we have that
  $\lambda^{A\alpha}_R P_A \Theta_\alpha= N^n_R=\lambda^\Gamma N_\gamma$ 
  and therefore
  $\lambda^{A\alpha}_R =\lambda^\Gamma k^{A\alpha}_\Gamma\,$. 
  Defining now
  \begin{align}
    \hat \omega^n &:= \bar\omega^n
      -\lambda^{\Delta\alpha}V_{\Delta\alpha}
      -\lambda^\Gamma W_\Gamma\;,\nonumber\\
    \hat 
    \omega^{n-1} &:=\bar\omega^{n-1}
    -\lambda^{\Delta\alpha}j_\Delta\Theta_\alpha
    -\lambda^\Gamma (B^{n-1}_\Gamma
    - k^{A\alpha}_\Gamma I^{n-1}_A\Theta_\alpha)\nonumber\\
    &~~= [P^{n-1\alpha}_R+dI^{n-2\alpha}_R
    +sK^{n-1\alpha}_R]\Theta_\alpha
    + B_R^{n-1} -\lambda^\Gamma B^{n-1}_\Gamma \;, 
   \nonumber
  \end{align}
  where in the last equality we used \eqref{C.60}, the cocycle
  relation $s\hat\omega^n+d\hat\omega^{n-1}=0\,$ is seen to be obeyed.
  As in the proof of (i) for lower form degree above, one then
  concludes that $\hat\omega^n\sim B^n+I^{n\alpha}\Theta_\alpha$ which
  gives the result.

\item[(ii)] Consider
  $\omega^{n}$ as in \eqref{eq:1}, so that 
  $s\omega^n+d\omega^{n-1}=0$ with
  $\omega^{n-1}=B^{n-1}+M^{n-1}+I^{n-1\alpha}_\mu\Theta_\alpha+B^{n-1}_\mu
  +\mu^{a\alpha}T_{a\alpha}+\mu^\gamma
  W_\gamma+\lambda^{\Delta\alpha}j_\Delta\Theta_\alpha +\lambda^\Gamma
  (B^{n-1}_\Gamma-k^{A\alpha}_\Gamma I^{n-1}_A\Theta_\alpha)$, or
  equivalently,
  $\omega^{n-1}=\mu^{a\alpha}T_{a\alpha}+\mu^\gamma W_\gamma+\hat
  B^{n-1}+\hat I^{n-1\alpha}\Theta_\alpha$, with
  $\hat B^{n-1}=B^{n-1}+B^{n-1}_\mu+\lambda^\Gamma B_\Gamma^{n-1}$,
  and
  $\hat
  I^{n-1\alpha}\Theta_\alpha=M^{n-1}+I^{n-1\alpha}_\mu\Theta_\alpha
  +\lambda^{\Delta\alpha}j_\Delta \Theta_\alpha -\lambda^\Gamma
  k^{A\alpha}_\Gamma I^{n-1\alpha}_{A}\Theta_\alpha$. The assumption
  that $\omega^n$ is trivial implies that so is $\omega^{n-1}$,
  $\omega^{n-1}\sim 0$.

  We have shown that (ii) holds at form degree $n-1$, so we get
  $\mu^{a\alpha}=0=\mu^\gamma$ which also implies
  $I^{n-1\alpha}_\mu=B^{n-1}_\mu=\hat B^{n}_\mu=b^n_\mu=0$,
  $B^{n-1}=-\lambda^\Gamma B_\Gamma^{n-1}$, and
  $\hat I^{n-1\alpha}\Theta_\alpha\approx
  N^{n-1}+dI^{n-2\alpha}\Theta_\alpha$. Since $N_\Gamma$ exists only
  if the form degree $n$ is even, there is no $N_\Gamma$, nor
  $B_\Gamma^{n-1}$ for $n$ odd, and thus $B^{n-1}=0=B^{n-1}_\Gamma=0$
  for $n$ odd.

  For $n$ even, $M^{n-1}=0$. We also have
  $d (\lambda^\Gamma B^{n-1}_\Gamma)+s(\lambda^\Gamma
  b^{n}_\Gamma)=\lambda^\Gamma N_\Gamma$. Using this, we find that
  $s (B^{n}+\lambda^\Gamma
  b^{n}_\Gamma)+d(M^{n-1}+B^{n-1}+\lambda^\Gamma
  B_\Gamma^{n-1})=\lambda^\Gamma N_\Gamma$. Since the second term on
  the left hand side vanishes, we get that $\lambda^\Gamma
  =0$. Indeed, $\lambda^\Gamma N_\Gamma$ is $s$ exact only if it
  vanishes. This shows that $B^{n-1}=0$ if $n$ is even as well.

  The equation for $\hat I^{n-1\alpha}$ with $\lambda^\Gamma=0$ now
  gives
  $\lambda^{\Delta\alpha}j_\Delta \Theta_\alpha \approx
  N^{n-1}+dI^{n-2\alpha}\Theta_\alpha$. Since
  $M^{n-1}-N^{n-1}=P^{n-1\alpha}\Theta_\alpha=dq^{n-2\alpha}\Theta_\alpha$,
  we find $\lambda^{\Delta\alpha}j_\Delta\approx d\omega^{n-2}$ which
  implies $\lambda^{a\Delta}=0$ since they represent non trivial
  characteristic cohomology classes. We then remain with
  $P^{n-1\alpha}\approx dI^{n-2\alpha}$. Using (iv) at $p=n-1$ then
  gives $P^{n-1\alpha}=0$ and then that $M^{n-1}=0=N^{n-1}$. It
  follows that $sB^{n}=0$ which implies $B^{n}=0$ since $B^{n}$ needs
  to have a non trivial descent. We remain with
  $I^{n\alpha}\Theta_\alpha=s\eta^{n}+d\eta^{n-1}$. This implies that
  $s\eta^{n-1}+d\eta^{n-2}=0$. Using (i) at $p=n-1$ we can assume that
  $\eta^{n-1}=\tilde B^{n-1}+\tilde
  I^{n-1\alpha}\Theta_\alpha+\rho^{a\alpha}T_{a\alpha}+\rho^\gamma
  W_\gamma+s\tilde\eta^{n-1}+d\tilde\eta^{n-2}$.  Using
  $d\tilde B^{n-1}=-s(\tilde B^n+b^n)+\tilde N^n$ and
  $d(\tilde I^{n-1\alpha}\Theta_\alpha)=d\tilde
  I^{n-1\alpha}\Theta_\alpha-s(\tilde I^{n-1\alpha}[\Theta_\alpha]^1)$
  gives
  $[I^{n\alpha}-d\tilde I^{n-1\alpha}-\tilde
  P^{n\alpha}-(\rho^{a\beta}(-)^nk_a^{n-2}+\rho^\gamma
  k_\gamma^{\bar\cA\beta})P^{2\alpha}_\beta]\Theta_\alpha=
  s(\eta^n-\tilde B^n-\tilde b^n-d\tilde \eta^{n-1}-\tilde
  I^{n-1\alpha}[\Theta_\alpha]^1-\rho^{a\alpha}U_{a\alpha}-\rho^\gamma
  R_\gamma)$ with $\tilde P^{n\alpha}\Theta_\alpha=\tilde N^n$,
  $P^{2\alpha}_\beta\Theta_\alpha=N^2_\beta$. This gives
  $I^{n\alpha}\approx d\tilde I^{n-1\alpha}+\tilde
  P^{n\alpha}+(\rho^{a\beta}(-)^nk_a^{n-2}+\rho^\gamma
  k_\gamma^{\bar\cA\beta})P^{2\alpha}_\beta$ and proves (ii) for
  $p=n$.

\item[(iii)] $I^n\approx d\omega^{n-1}$ implies $I^n\sim 0$. Using (ii)
  at $p=n$, which we have already proved, gives
$I^n\approx P^n+dI^{n-1}$ since when $\Theta_\alpha$ reduces to $1$,
$N^2_\alpha=0$. The converse is again trivial since $P^n=dq^{n-1}$. 

\end{enumerate}

\addcontentsline{toc}{section}{References}

\providecommand{\href}[2]{#2}\begingroup\raggedright\endgroup

%\bibliography{C:/Users/gbarn/Dropbox/Literature/master}
%\bibliographystyle{C:/Users/gbarn/Dropbox/Literature/utphys}
%\bibliography{/Users/nicolasboulanger/Dropbox/Biblio}
%\bibliographystyle{/Users/nicolasboulanger/Dropbox/utphys}

\end{document}